\documentclass[12pt]{article}%
\usepackage{graphicx}
\usepackage{amsmath}
\usepackage{amsfonts}
\usepackage{amssymb}%
\providecommand{\U}[1]{\protect\rule{.1in}{.1in}}
\usepackage{color}

\begin{document}

\title{Retirement Spending and Biological Age}
\author{H. Huang, \; M. A. Milevsky\thanks{Milevsky (the contact author) is professor of finance at the Schulich School of Business and can
be reached via email at: milevsky@yorku.ca, or at Tel: 416-736-2100 x 66014.
His mailing address is: 4700 Keele Street, Toronto, Ontario, Canada, M3J 1P3.
Huang is a professor of mathematics at the Department of Mathematics and
Statistics, York University and Deputy Director of the Fields Institute in Toronto, hhuang@mathstat.yorku.ca. Salisbury is a professor of mathematics at the Department of Mathematics and Statistics, York University, salt@mathstat.yorku.ca. The authors would like to acknowledge helpful comments from David Blake, Melanie Cao, Helmut Gruendl, Steve Haberman, Raimond Maurer, David Promislow, Pauline Shum, Yisong Tian, seminar participants at Goethe University, York University, ARIA, detailed comments from an associate editor and reviewer at JEDC, as well as funding from NSERC (Salisbury and Huang), the IFID Centre and a Schulich Research Fellowship (Milevsky).} \; and T. S. Salisbury}
\date{11 September 2017}
\maketitle



\begin{abstract}

We solve a lifecycle model in which the consumer's {\bf chronological} age does not move in lockstep with calendar time. Instead, {\bf biological} age increases at a stochastic non-linear rate in time like a broken clock that might occasionally move backwards. In other words, biological age could actually decline. Our paper is inspired by the growing body of medical literature that has identified biomarkers which indicate how people age at different rates. This offers better estimates of expected remaining lifetime and future mortality rates. It isn't farfetched to argue that in the not-too-distant future personal age will be more closely associated with biological vs. calendar age. Thus, after introducing our stochastic mortality model we derive optimal consumption rates in a classic Yaari (1965) framework adjusted to our proper clock time. In addition to the {\em normative} implications of having access to biological age, our {\em positive} objective is to partially explain the cross-sectional heterogeneity in retirement spending rates at any given chronological age. In sum, we argue that neither biological nor chronological age alone is a sufficient statistic for making economic decisions. Rather, {\bf both} ages are required to behave rationally. 

\end{abstract}

\clearpage

\section{Introduction and Motivation}

In the classical Yaari (1965) model -- and the vast number of 
{\em lifecycle + mortality} papers it has spawned over the last five decades, such as Levhari and Mirman (1977), Davies (1981), or Feigenbaum (2008a)  -- the operating assumption is that the chronological age of the representative consumer is the only time variable that matters. In these {\em deterministic mortality} models there is a known and consistent mapping between age, time and the future hazard rate, all of which then flows into preferences via discounted utility. 

In this paper we take the first steps towards solving a lifecycle model in which the consumer's age does not move in lockstep with calendar time. Instead, biological age increases at a stochastic non-linear rate in time which is a clock that might occasionally move backwards. Indeed, many papers in the economics literature have examined the implications of uncertainty in interest rates, investment returns, wages and income on optimal consumption and precautionary savings\footnote{See for example Feigenbaum (2008b) for a discussion and analysis of the impact of uncertainty in future income on savings rates.}. This paper examines the implications of uncertainty in time (or age) itself on optimal consumption and spending. In addition to trying to understand the {\em normative} implications, or how one might use this (new) information in a lifecycle model, our {\em positive} empirical objective is to shed new light on the cross-sectional heterogeneity in retirement spending rates. As of late this has been somewhat of a puzzle for researchers in the retirement and pensions arena.

Indeed, one of the stylized facts within the empirical lifecycle literature is the wide dispersion or heterogeneity of consumption and/or withdrawal rates during the period commonly referred to as retirement. At any given retirement age these rates vary cross-sectionally even when controlling for financial wealth, pensions and other economic variables. To be clear, when we use the term {\em withdrawal rate} in this paper we mean the amount of dollars extracted or removed from investible net-worth in any given year, presumably for consumption purposes. When we use the term retirement {\em spending rate} we are referring to the same withdrawal amount but expressed as a percentage of financial net worth at that point in time.\footnote{So, a withdrawal rate of $\$50,000$ per year at age 65 from a (total) net-worth of $\$1,000,000$ at age 65, is labeled a spending rate of $5\%$ at age 65. Ten years later the same 75 year-old might continue to withdraw at the rate of $\$50,000$ per year from his or her portfolio, but if the value of the portfolio has declined to $\$500,000$ (for example) the spending rate at age 75 is (obviously) now $10\%$. Likewise, as far as terminology is concerned, a retiree with (government or corporate) pension income would add that amount to their withdrawals to arrive at total consumption. In other words, retirement consumption, withdrawal (occasionally called draw-down in the UK) and spending rates are all different metrics.  And, of course, there is Gary Becker's well-known distinction between expenditure and consumption. Now there is some confusion (especially among practitioners) over terminology which is why we want to get this issue out of the way.} 

\begin{figure}
\caption{{\em {\bf Source:} Statistics Canada, Survey of Financial Security 2012. Market value of 10 financial categories + 5 non-financial categories minus the value of 7 debt categories.}}
\begin{center}
\vspace{-0.5in}
\includegraphics[width=0.75\textwidth]{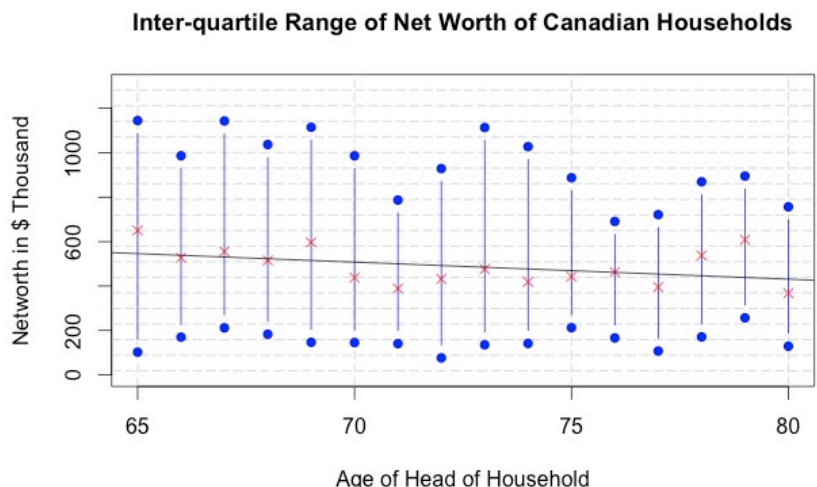} 
\label{A}
\end{center}
\vspace{-1.0in}
\end{figure}

To our main point, as far as the data is concerned retirement withdrawal rates (\$) and/or spending rates (\%) are quite heterogenous. As just one example, Figure~\ref{A} displays the economic net-worth of households in Canada, in which the head of the household is between the age of 65 and 80. The vertical bars capture the inter-quartile range and the midpoint is the mean (average). The data displayed are from the most recent Survey of Financial Security (SFS), which is collected and made available every 5 years. For example, the average household at the age of 65 had a net worth of \$660,000. At age 80 the average is approximately \$380,000. Regardless of the actual numbers, Figure~\ref{A} illustrates a general trend for how wealth evolves during retirement {\em anywhere} in the world and not only in Canada. It declines on average but with substantial heterogeneity.

And, even though the SFS data is a snapshot of the entire population of Canada at a point in time -- and doesn't address retirement withdrawals of a given cohort {\em per se} -- it sets the stage for our discussion. In a simple linear regression with age $x_i$ as the {\em independent} variable and economic net-worth $w_i$ as the {\em dependent} variable: $w_i \; = \; \alpha_0 \; + \; \alpha_1 \, x_i \; + \; e_i$, the value of net-worth declines by approximately $\alpha_1=-\$18,000$ per (chronological age) year. That would correspond with a population average {\em withdrawal rate.} That number is for the 3,179 sampled households above the age of 65 in the SFS2012 dataset, which is the most recent wave of the the survey. If we limit ourselves to financial assets (a.k.a. investment accounts) and scale withdrawals by the value of the assets, they decline by approximately $4\%$ per year. That number would represent the population average {\em spending rate}\footnote{Strictly speaking we should adjust our figures to account for asset returns. The point here is the trend.} and is uncannily close to the widely used (among financial planners) 4\% rule, proposed by William Bengen in the early 1990s. 

The above evidence or behavior is (loosely) consistent with the Modigliani life-cycle model in which consumers accumulate wealth while working and slowly drawdown to consume during their retirement years. The question of whether or not the average or representative consumer is depleting their wealth {\em fast enough} is a fertile area of research in the academic literature, but is not our focus. 

Notwithstanding the trend line, it is interesting to note the wide dispersion in spending rates for retirees who are of the same chronological age, even after controlling for the level of wealth and asset holdings. For example (within the above-mentioned SFS data set) at the age of 70 over 1/4 of Canadian households have a negative portfolio withdrawal rate (i.e. they continue to save).  At the age of 70 the median retirement spending rate (relative to wealth one year later at age 71) is 10.48\%. Of course not all 70-year old households spend at the rate of 10.48\%. According to the SFS one quarter of this group spends less than 4.35\% and at the other extreme another quarter experienced an increased of 7.45\% (or more) in their net worth. Again, a non-trivial group is not drawing-down wealth while others spend (much more) than the 4\% rule would dictate. The dispersion in withdrawal or spending rates is not unique to Canada. See for example, Coile and Milligan (2009) in the U.S. Similar evidence is provided by Poterba, Venti and Wise (2015), as well as De Nardi, French and Jones (2015), or Banks, Blundell, Levell and Smith (2015). In this regards, see also Love, Palumbo and Smith (2009) which examines the trajectory of wealth itself in retirement. All of this is serves as motivation. Within a lifecycle framework, what might explain the reason some retirees (at the same age) withdraw or spend (much) faster than others?  Broadly speaking the literature offers three categories of explanations for the dispersion in these rates.

\subsection{Why the Dispersion in Spending Rates?}

\begin{enumerate}

\item Heterogeneity in {\bf Leisure, Labor \& Legacy Preferences:} See for example the work by Gan, Gong, Hurd and McFadden (2015), Hubener, Maurer and Mitchell (2015), as well as Farhi and Panageas (2017). Individuals with weaker bequest motives, or greater potential for future labour income or lower utility of leisure might all be inclined to spend more today relative to the average retiree. The heterogeneity of preferences would induce statistical dispersion in observed withdrawal and spending rates. 

\item Heterogeneity in {\bf Portfolio Choice, Markets \& Investment Views:} See for example the work by Cocco, Gomes and Maenhout (2005), Horneff, Maurer, Mitchell and Stamos (2009), Cocco and Gomes (2012), Yogo (2016). Investors who perceive (or believe) that their portfolio or individual investments will earn greater returns are likely to spend or withdraw more, all else being equal. This would also tie into the behavioral finance and economics literature on irrational beliefs. 

\item Heterogeneity in {\bf Longevity \& Mortality Expectations.} See for example Groneck, Ludwig and Zimper (2016), Kuhn, Wrzaczek, Prskawetz, Feichtinger (2015), or Spaenjers and Spira (2015). While population mortality rates are measurable and objective, individuals might have personal (perhaps incorrectly estimated) views of their survival probabilities, which would induce them to spend more or less than the average individual.  This idea or explanation can also be positioned within the context of health heterogeneity, see for example the work by Rosen and Wu (2004), Berkowitz and Qiu (2006) or De Nardi, French and Jones (2010). 

\end{enumerate}

Our current work falls within category \#3 and more specifically in the measurement of (subjective) mortality, but we frame and formulate the lifecycle problem quite differently. Instead of splitting the universe into five health categories for example -- and rhetorically asking people to determine whether they are in {\em very good} health or merely in {\em good health} -- we query or carefully measure their biological age. Once again we are interested in examining the impact of uncertainty in future age itself on optimal consumption, similar in spirit to prior work in economics that has examined the impact of uncertainty in wages, income and returns on precautionary saving.

Generally speaking, we assume a canonical retiree with zero weight on bequest motives (eliminating explanation \#1) and a portfolio that only consists of a risk-free asset (eliminating explanation \#2) to focus attention on the {\em definition of aging} and its impact on portfolio withdrawal and spending rates. At the risk of placing the cart before the horse, we find that the dispersion or heterogeneity of biological ages -- at any given chronological age -- can (partially) explain the heterogeneity of spending rates. Our (main) contribution is to illustrate how to embed this within a rational lifecycle model.

\subsection{How Old Are You, Really?}

There is a growing body of medical evidence (and mail-order kits) suggesting that an individual's true age can be measured (more accurately) using telomere length. Telomeres are the protective ends of chromosomes and (the claim is that) their length provides the best biomarker of aging. Moreover, {\em biological age} can diverge by as much as 10 to 15 years from chronological age as measured by calendar years. In other words, a 65 year-old retiree might in fact be as young as 50 or as old as 80 when measured properly in terms of forward-looking mortality and morbidity rates.  And, while the technology to accurately estimate biological age is being refined -- and other biomarkers of aging might emerge -- this does raise the possibility that individuals will soon have access to a {\em new age} for retirement. It's not inconceivable that chronological age will take a back seat to biological age in the public discourse around retirement policy.

Now, biological age should not be viewed simply as an age set-back on a (deterministic) mortality table or a fixed scaling adjustment factor, both of which are quite common in actuarial practice and the health economics literature. Indeed, the joint dynamics of biological (B) age and chronological (C) age are subtle and mathematically non-trivial. For example, the divergence between B-age and C-age is highest around middle age and lowest at younger and older ages. Intuitively, a (live) centenarian's B-age is quite close to her C-age and vice versa. In other words the dispersion of relative ages within a population is (chronologically) age dependent. For example, in one (very) widely cited socioeconomic study by Marmot and Shipley (1996), the dispersion in mortality rates -- or what we would call biological age -- continues well-into retirement\footnote{For example, they found that a retiree who had worked as an administrative clerk had a 50\% higher mortality rate at the age of 90, compared to someone who has worked as an executive within the same governmental organization.}.

Motivated and inspired by this new view of ageing, in this paper we solve a classical lifecycle model of consumption and spending in which the (rational) economic agent has two distinct and measurable ages at every point in time; which we call {\it B-age} and {\it C-age}. We assume a canonical retiree with a fixed endowment of investable wealth and then derive the optimal consumption rate as a function of the two-dimensional age co-ordinates. 

In the absence of any pension income, the consumption rate (in \$) is the withdrawal rate (in \$) and the spending rate we defined earlier is the ratio of consumption to wealth and is expressed as a percent (\%). Of course, our framework collapses to the standard lifecycle model and the known Yaari (1965) results when B-age is forced to equal C-age at all times.

Probabilistically, we connect the two distinct ages via the mechanism of stochastic mortality rates. Namely, by formulating a generalized {\em Brownian Bridge}-driven model for individual mortality rates and then inverting the standard (population) Gompertz law of mortality to arrive at one's relative B-age. All of this will be carefully explained but at this point we should note that introducing or using the Brownian Bridge as our stochastic mortality model is quite novel (in our opinion) and has not been proposed before in the actuarial arena and certainly has not been put to work in the lifecycle economics literature. We will address why a Brownian Bridge is a suitable model for (stochastic) mortality -- as well as the basic question of {\em what is} a Brownian Bridge?

\subsection{Connection to the Actuarial Literature \label{LIT}}

Our paper sits at the intersection of three different fields or areas of research; (i.) the economics literature and the data which motivated the paper belong firmly in the lifecycle arena, (ii.) bio-gerontology and the study of aging, and (iii.) actuarial science and demography.

Regardless of the field -- economics, actuarial science or gerontology -- a very popular assumption regarding death and aging is that an individual's hazard rate at age $x$ (denoted by $\lambda_x^{\text{G}}$) obeys the so-called Gompertz law of mortality, namely $\lambda_x^{\text{G}}=\frac{1}{b}\exp \left( \frac{x-m}{b}\right)$. 
The way this law is expressed, $x$ is current age, the parameter $m$ is a modal value of life in years (e.g. 80) and $b$ is a dispersion parameter in years (e.g. 10). Stated briefly, mortality rates grow predictably and deterministically by approximately 9\% to 10\% per year. Most of the lifecycle papers in the financial and economics literature explicitly or implicitly assume this law of mortality on the individual level, perhaps calibrated to discrete population mortality tables. For example, Leung (2007) who extended and refined the original Yaari (1965) model, used this exact law of mortality in his numerical examples.

The Gompertz (or the variant known as Gompertz-Makeham) law is the canonical and prototypical {\em deterministic} model of mortality. It is widely taught and used in pricing insurance and annuity contracts, albeit after some adjustments for anti-selection and various discretization and smoothing techniques. 

The first extension of the (deterministic) Gompertz law to a stochastic environment was the work by Lee \& Carter (1992). From that starting point the demographic and actuarial literature have proposed many models for the evolution of population mortality rates in which the hazard rate is assumed to be a diffusion process in continuous time. In this matter we refer readers to the actuarial papers by Milevsky \& Promislow (2001), Dahl (2004), Biffis (2005), Renshaw \& Haberman (2006), Cairns, Blake and Dowd (2006), Schrager (2006), Luciano \& Vigna (2008), Plat (2009), Cairns, et. al. (2011), Huang, Milevsky and Salisbury (2012), Blackburn and Sherris (2013), Delong \& Chen (2016), Liu \& Lin (2012). See also Pitacco, Denuit, Haberman \& Olivieri (2008) as well as the criticism by Norberg (2010). 

To be clear in our positioning, this paper is not an attempt to better forecast or project future population mortality rates. Rather, our assumption is that today's 65 year-old does not know (with certainty) what his or her mortality rate will be in 30 years. They only have a rational expectation. Our question is how the uncertainty (and the ability to adjust later once that uncertainty abates) affects optimal behavior. Our language (biological vs. chronological age) is borrowed from the literature on bio-demographics. We refer the interested readers to the work by Cawthon et al. (2003), Dong, Milholland \& Vijg (2016), Heidenger, et. al. (2012), Mather, et. al. (2010) and in particular Olshansky, Carnes \& Cassel (1990) for a general discussion of the uncertainty in future mortality rates and its relation to {\em the limits to life.}
 
This paper merges (i.) lifecycle models and (ii.) stochastic mortality on the individual level. Aside from the attempt to explain {\em current} and observed behavior, we develop a normative model for the future when the measurement of biological age will presumably be much more accurate. Once everybody knows their B-age, how will consumption respond?

\subsection{Overview of the Paper}

The remainder of this paper is organized as follows: Section~\ref{STO} describes and explains our stochastic process for the instantaneous mortality rate -- the so called Brownian Bridge model -- which then leads to the distinct evolution of biological and chronological age. In section~\ref{LCM} we present the theoretical core of the paper, which is the derivation of the optimal consumption and withdrawal rate as a function of both biological and chronological age. In that section we carefully explain the difference between a model with {\em deterministic} aging, which has been part of the lifecycle literature for decades and our model of {\em stochastic} aging. Then in section ~\ref{NUM} we provide a range of numerical results as well as the testable implications emerging from our framework. Finally, section~\ref{CON} concludes the paper. The technical proofs and mathematical derivations are relegated to an appendix.

\section{Stochastic Model of Aging \label{STO}}
\subsection{How to Think About Death}

As far as the modeling is concerned we assume that one can accurately measure (e.g. using a combination of telomere length, systolic blood pressure, body-mass index, etc.) an individual's current mortality rate and that it can be higher or lower than (average) population mortality rates at that chronological age.  This then allows us to invert a (population) Gompertz law using the observed mortality rate to obtain the corresponding biological age. Stated differently, our first assumption is that the two-century old Gompertz law of mortality applies in (something we call) biological time rather than in standard calendar time.

Second, we assume a convenient functional form in which, at some fixed age (e.g. $110$) the mortality rate of our canonical retiree will reach $\lambda_T=1$ with probability one and that life expectancy from that point onward is exactly one year. You stop aging. 

\begin{figure}
\caption{{\em Example of a 90\% confidence interval (band) for the evolution of biological age, assuming that B-age and C-age agree at 60 and 110. We take $\lambda_{60}=0.005,\lambda_{110}=1.0$, $\xi=1$ and volatility $\sigma=0.60$}}
\vspace{-0.1in}
\begin{center}
\includegraphics[width=0.7\textwidth]{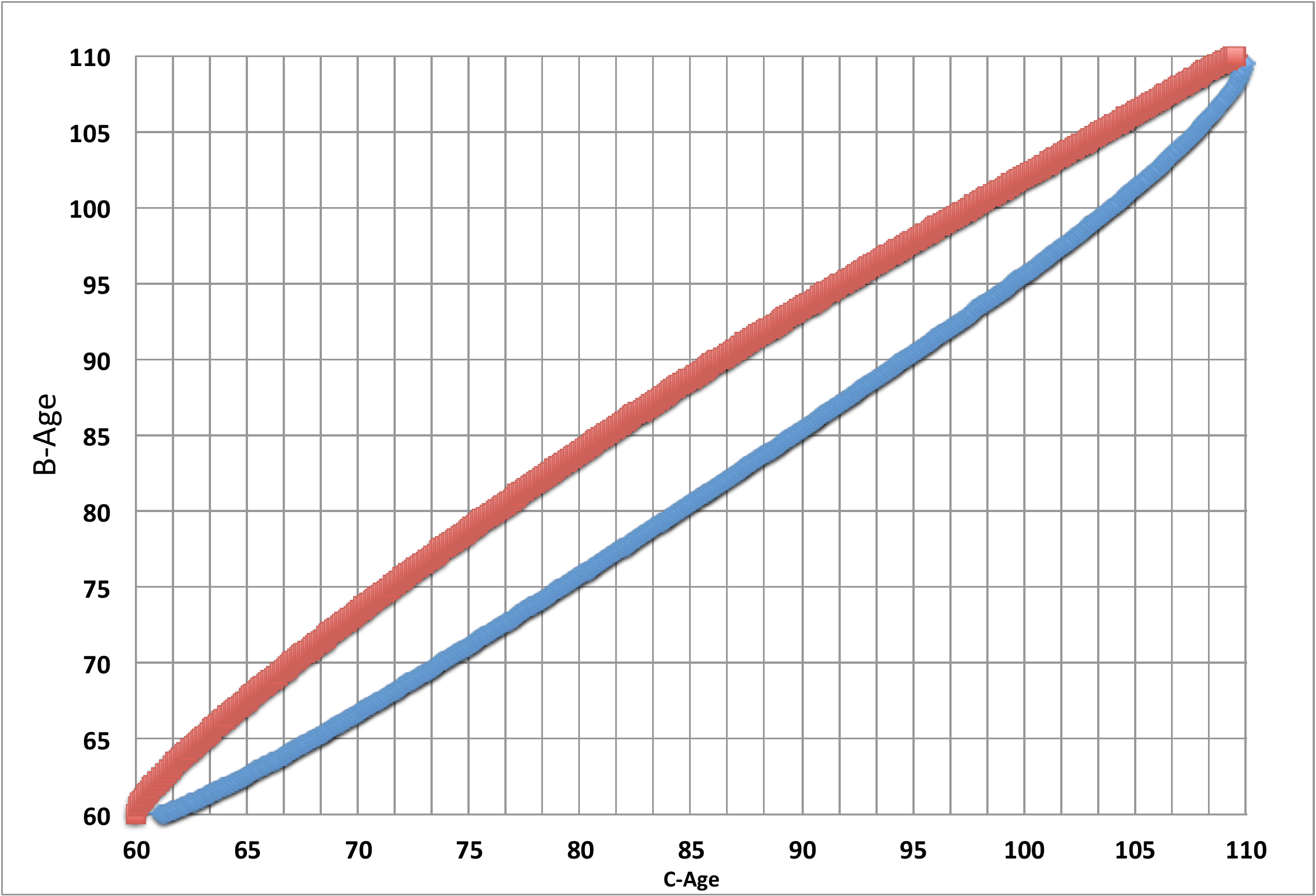} 
\label{BvC}
\end{center}
\vspace{-0.3in}
\end{figure}

Third and finally, we assume that the instantaneous mortality rate which follows a diffusion process {\em wanders} randomly over time but reverts back to the above mentioned $\lambda_T=1$ by time $T$ and is {\em absorbed} at that end point. Figure~\ref{BvC} offers a picture of the evolution of the two ages with a corresponding 90\% confidence interval assuming both ages are identical at time zero (which here is age 60). There are a number of parameter assumptions ($\xi, \sigma$) that have been used to generate the figure, all of which will be explained later.

\subsection{Process for the Mortality Rate}

Let $A_t$ denote biological age and $\kappa_t$ chronological age, where $\kappa_t=t+\kappa_{0}$, for some offset $\kappa_{0}$ and therefore $d\kappa_t=dt$. Also, $A_T=\kappa_T$, for a fixed value of $T$, and $A_t=A_T$ for $t\ge T$. We will typically assume that $A_0=\kappa_0$ (and will note occasions where this assumption is dropped). We assume an individual Gompertz mortality law based on biological age, so the hazard rate is $\lambda_t=\frac1b e^{\frac{A_t-m}{b}}$, using the common $(m,b)$ parameters. Alternatively, for the formulation we prefer and use in this paper, we can express this in terms of the corresponding (pinned down) hazard rates $\lambda_0$ and $\lambda_T$ at times $0$ and $T$ (rather than $m$ and $b$, common in the actuarial arena), namely:

\begin{equation*}
\lambda_t=\frac{1}{b}e^{\frac{A_t-m}{b}}=\lambda_0e^{\frac{A_t-A_0}{b}}=\lambda_0 e^{\frac{A_T-A_0}{b}\cdot \frac{A_t-A_0}{A_T-A_0}}=\lambda_0 \Big(\frac{\lambda_T}{\lambda_0}\Big)^{\frac{A_t-A_0}{A_T-A_0}}.
\end{equation*}
The two sets of parameters $(\lambda_0,\lambda_T)$ and $(m,b)$ are interchangeable, with
\begin{equation}
b=\frac{\kappa_T-\kappa_0}{\log(\lambda_T/\lambda_0)}, \hspace{0.5in} 
m=\kappa_0-b\log(b\lambda_0)
\label{mb}
\end{equation} 

To keep things simple and avoid clutter, the mortality hazard rate will be expressed in terms of a biological-age variable $a$ via a function $\lambda(a)$, which is written and expressed as:\footnote{Just to be clear when it comes to calibration, one can assume that the end-points  $(\lambda_0,\lambda_T)$ are known and then solve for the implied Gompertz parameters $(m,b)$. Or, {\em vice versa} one can start with a particular Gompertz parameter set and then compute the relevant end-points. For example, if we assume that mortality is pinned at $\lambda_0=0.005$ at the age of $x=60$ and $\lambda_{50}=1$ at the age of $x=110$, then the implied Gompertz values are $m=88.8174$ and $b=9.4369$, according to equation~(\ref{mb}).}
\begin{equation*}
\lambda(a)=\lambda_0\Big(\frac{\lambda_T}{\lambda_0}\Big)^{\frac{a-\kappa_0}{\kappa_T-\kappa_0}}.
\end{equation*}

We assume that the biological age can be written as $A_t=\kappa_t + Y_t$ for $t\le T$, where 
\begin{equation}
dY_t=-\xi\frac{Y_t}{T-t}\,dt + \sigma\,dB_t.
\label{dynamicsforY}
\end{equation}
Since $d\kappa_t=dt$ this implies that, for $t<T$,
\begin{equation}
dA_t=\Big(1+\xi\frac{\kappa_t-A_t}{T-t}\Big)\,dt + \sigma dB_t.
\label{dynamicsforA}
\end{equation}
In other words, $\xi$ is a type of mean reversion parameter and $\sigma$ a volatility parameter. The singularity of the drift at $t=T$ will force $A_T=\kappa_T$, as we will show below. If $\xi=\sigma=1$ then $Y_t$ is in fact a well-known stochastic process called a {\it Brownian Bridge}, which arises in studies of the conditional distributions of Brownian motion. It has the property that it is pinned to take value 0 at both $t=0$ and $t=T$, which we will now show carries over to $Y_t$ under other parameter values. This is the reason we will refer to either $Y_t$ or $A_t$ as a {\it generalized Brownian Bridge} process.

By equation \eqref{dynamicsforY}, we have that $d[(T-t)^{-\xi}Y_t]=\sigma(T-t)^{-\xi}\,dB_t$, 
so: 
\begin{equation*}
(T-t)^{-\xi}Y_t-(T-s)^{-\xi}Y_s=\sigma \int_s^t(T-q)^{-\xi}\,dB_q. 
\end{equation*}
In other words, given the history of biological age until time $t$, $Y_t$ has a normal distribution with conditional mean $(\frac{T-t}{T-s})^\xi Y_s$ and conditional variance $\sigma^2(T-t)^{2\xi}\int_s^t(T-q)^{-2\xi}\,dq$. If $\xi\neq\frac12$ this $=\sigma^2\frac{T-t}{2\xi-1}[1-(\frac{T-t}{T-s})^{2\xi-1}]$. In particular, the mean and variance of $Y_t$ both $\to 0$ as $t\uparrow T$, so $Y_t\to 0$ in probability as well. One can show that $Y_t\to 0$ a.s. 

It's important to emphasize that while the distribution of $Y_t$ is symmetric around zero based on the construction of equation \eqref{dynamicsforY}, one is {\bf not} likely to observe a chronological $\kappa_t=$ 75-year old with a biological age of $A_t=95$ and $A_t=55$ with equal odds. Indeed, the path above zero for which $Y_t>0$ is more hazardous and is more likely to kill the retiree along the human lifecycle relative to the path under zero for which $Y_t<0$. In other words, although the odds of both events are small (and technically have measure zero), it is more likely one will observe the co-ordinates $(\text{B-age}=55,\text{C-age}=75)$ vs. $(\text{B-age}=95,\text{C-age}=75)$. We will return to these odds later on.

\subsection{Intuition for the Stochastic Model}

Given that our model for mortality (life and death of the retiree) is quite different from the standard models used in the lifecycle economics literature, in this subsection we provide some additional intuition and insight into the difference between stochastic and deterministic aging. 

\begin{table}
\caption{{\em 90\% Confidence Interval for Biological (B) Age at Chronological (C) Age 85 (time $t=25$), under various combinations of reversion speed $\xi$ and mortality volatility $\sigma$. }} 
\label{agerange}
\begin{center}
    \begin{tabular}{||c|c|c|c||} \hline
        \multicolumn{4}{||c||}{{\bf Lower Endpoint, Upper Endpoint \& Range}} \\ \hline
        ~ & $\sigma=0.30$ &  $\sigma=0.60$ &  $\sigma=0.90$ \\ \hline    
        $\xi=0.50$ & $[81.33, 88.17] \;\;\; 6.84 \, yrs$ &  $[79.67, 89.42] \;\;\; 9.75 \, yrs$ & $[77.67, 90.67] \;\;\; 13.00 \, yrs$ 
        \\ \hline
        $\xi=0.75$ & $[81.67, 87.92] \;\;\; 6.25 \, yrs$ &  $[80.08, 89.08] \;\;\; 9.00 \, yrs$ & $[78.25, 90.33] \;\;\; 12.08 \, yrs$ 
        \\ \hline
        $\xi=1.00$ & $[81.92, 87.67] \;\;\; 5.75 \, yrs$ &  $[80.50, 88.75] \;\;\; 8.25 \, yrs$ & $[78.75, 90.00] \;\;\;11.25 \, yrs$ 
        \\ \hline
        $\xi=2.00$ & $[82.58, 87.08] \;\;\; 4.50 \, yrs$ &  $[81.50, 87.92] \;\;\; 6.42 \, yrs$ & $[80.25, 88.92] \;\;\; 8.67 \, yrs$       
       \\ \hline
       \multicolumn{4}{||c||}{Assumes $\lambda_{0}=0.005$ when B-age=60=C-age, \& $\lambda_{50}=1.00$ when B-age=110=C-age. }
        \\ \hline \hline
        
\end{tabular}
\end{center}
\vspace{-0.2in}
\end{table}

Table \ref{agerange} offers some numerical values for the 90\% confidence interval of biological age when the retiree is 85 years old, chronologically. It focuses on the impact of the two key parameters $\xi$ and $\sigma$ in our model. Notice that as the volatility of mortality $\sigma$ increases from $0.30$ to $0.90$, the range of possible biological age increases. For example when $\xi=1$, which is the canonical basis for most of our numerical examples to follow, at the chronological age of 85 (which is also time $t=25$), biological age can vary by 5.75 years when $\sigma=0.30$ and by as much as 11.25 years when $\sigma=0.90$. This effect is quite natural and to be expected given the definition of volatility which is synonymous with dispersion. The same outcome can be observed when the speed or force of mean reversion $\xi$ is reduced from 1.0 to 0.50. The dispersion in range increases from 5.75 years to 6.84 years. Intuitively, the underlying diffusion process is wandering more (i.e. not forced to ``return quickly'') in between the fixed end-points.

These represent $90\%$ confidence intervals, which to be precise means there is a $5\%$ probability of observing a B-age {\em below} the lower bound and a $5\%$ probability of observing a B-age {\em above} the upper bound at the fixed Chronological age of 85. Obviously, using $95\%$ or $99\%$ values would increase the range of plausible B-ages. 

On a more subtle level it is worth examining the upper and lower bounds themselves relative to age of 85. Notice that all 12 ranges displayed are left-skewed. This is not coincidental or the result of numerical approximations. For example, in our canonical case of $\xi=1.0$ and $\sigma=0.30$ the lower bound is biological age 81.92 which is 3.08 years below chronological age 85. The upper bound is biological 87.67, which is only 2.67 years above 85. Indeed, the two numbers add-up to the listed range of 5.75 years. So even though $A_0=\kappa_0$ at time zero, as time evolves a lower B-age is more likely to be observed (among survivors) than a higher B-age.  The asymmetry is driven by the fact that {\bf if} biological age happens to wander above chronological age over time, the higher mortality rate is more likely to kill the retiree. He/she is less likely to survive to C-age 85. In contrast, if biological age wanders and remains under chronological age, the implied mortality rate is lower (by definition) and one is more likely to survive to C-age 85. This generates the skewness observed in the table.

In other words although the underlying stochastic process for the (generalized) Brownian bridge $Y_t$ is perfectly symmetric around zero and the unconditional biological age process $A_t$ evolves symmetrically around $\kappa_{t}$, once we translate into mortality units and then condition on survival the symmetry is destroyed. Only those individuals who survive till time $t$ contribute to the distributions displayed here. For example, in the upper right corner of the table in which $\xi=0.5$ and $\sigma=90\%$ the lower bound is 77.67 years (7.33 years under 85) and the upper bound is 90.67 (only 5.67 years above 85.) Very loosely speaking, this is a $55\%$ chance of being younger than your age and a $45\%$ chance of being older even though you started out being {\em exactly} your age. 
\begin{figure}[t!]
\caption{{\em The PDF of your Biological Age assuming you are alive at age 35, 60 or 85.}}
\label{3densities}
\vspace{-0.2in}
\begin{center}
\includegraphics[width=0.8\textwidth]{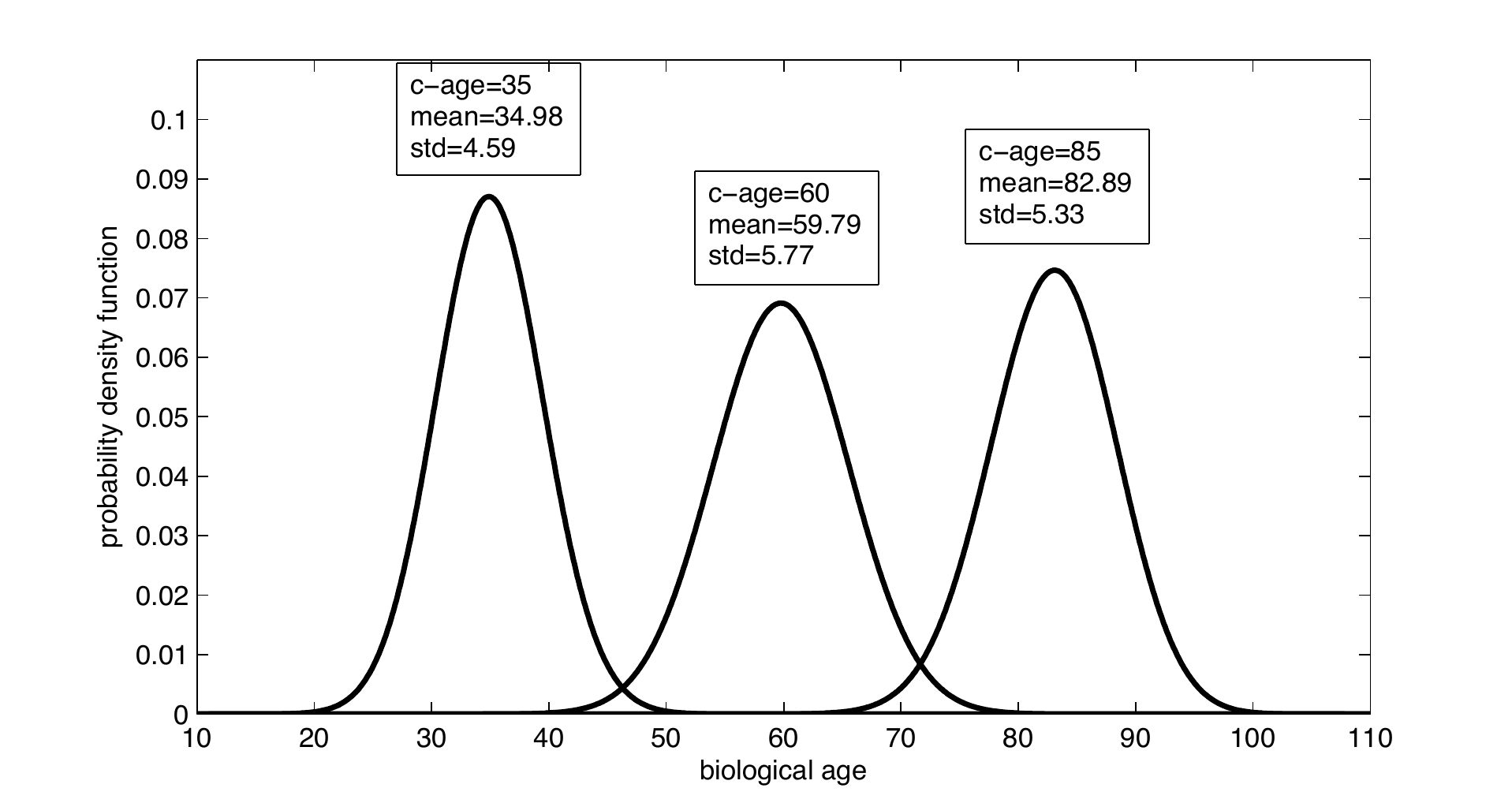} 
\end{center}
\vspace{-0.3in}
\begin{center}
{\em Parameters are $\xi=1$, $\sigma=0.30$, with $\lambda_{0}=0.0005$ when \\ B-age=10=C-age, and $\lambda_{100}=0.5$ when B-age=110=C-age.}
\end{center}
\end{figure}

To provide one final perspective before we move on to moments and economics, Figure \ref{3densities} displays the entire probability distribution of biological age at chronological ages 35, 60 and 85. In this particular figure we pin-down mortality rates at age 10 and at age 110, which means that they wander for 100 (chronological) years before converging at the end of the mortality table. The main graphical insight is the same as in Table \ref{agerange}, though Figure \ref{3densities} uses only one choice of $\xi$ and $\sigma$. A (live) chronological 85 year-old could conceivably have a (biological) mortality rate that is closer to that of a 70 year old, although the probability of observing this 15 year gap in ages is quite rare. The left tail is thin. But those (small) odds are relatively higher than the odds of observing a chronological 85 year-old who is 100 biologically. On the right side of that particular density curve the tail value is essentially zero. Stated differently, the large hazard rate would have surely killed the 85 year-old. This is why the three displayed mean values for B-age are always lower than C-age. 

We pause here to stress (yet again) that this is a fundamental aspect of our model for stochastic mortality; a twisted bridge.

\subsection{Expected Remaining Lifetime}
\label{sec:means}

Let $e(t,a)$ denote the life expectancy of an individual whose biological age at time $t$ is $a$. If $\zeta$ denotes our individual's lifetime then on the event that $\zeta>t$ we have 
\begin{equation*}
t+e(t,A_t)=E[\zeta\mid\mathcal{F}_t].
\end{equation*}
where $\mathcal{F}_t$ represents the historical information set at time $t$. The latter is a martingale, because of the Law of Iterated Expectations.  It also has a jump of $-e$ at $t=\zeta$. We may create a new martingale $N_t$ by starting with the jump process $-e(\tau,A_\tau)1_{\{t<\zeta\}}$ and then subtracting the integral of the jump intensity times the jump distribution. This integral is called the {\it compensator} of the jump and is introduced to help (allow) us to obtain expressions for moments and expectations. It follows that $E[\zeta\mid\mathcal{F}_t]-N_t$ is a continuous martingale.  Therefore
\begin{equation}
H_t=t+e(t,A_t)-\int_0^t \lambda_s e(s,A_s)\,ds
\label{eqn:compensatedprocess}
\end{equation}
is a continuous martingale when stopped at $\zeta$, so has drift $=0$. We may computing $dH_t$ via It\^o's lemma, and setting the drift to 0 now gives that
\begin{equation}
1+e_t(t,a) + \Big(1+\xi\frac{\kappa_t-a}{T-t}\Big)e_a(t,a) +\frac{\sigma^2}{2}e_{aa}(t,a)-\lambda(a) e(t,a)=0
\label{meanlifetimeeqn}
\end{equation}
for $t<T$, with boundary conditions $e(T,\cdot)=\frac{1}{\lambda_T}$ and $e(t,\infty)=0$. 
For the boundary condition at $a=-\infty$, observe from the above mean-variance calculation that if $s<t<T$ then $A_s\ll 0\Rightarrow A_t\ll0$.  Therefore when $a\to-\infty$, we will have $\lambda\to 0$. In other words we live to time $T$ and then die exponentially: $e(t,-\infty)=T-t+\frac{1}{\lambda_T}$. 

\begin{table}
\caption{{\em Expected Remaining Lifetimes. Mortality rate wanders as a Brownian Bridge with $\xi=1$ and $\sigma=30\%$. Assumes a terminal (age 110) mortality rate of $\lambda_{50}=1$. Gompertz parameters $b$ and $m$ are chosen so that at C-age 60, if B-age also $=60$ then $\lambda_0=0.005$}}
\label{erl}
\begin{center}
    \begin{tabular}{|c|c|c|c|c|c|c|c|c|}
        \hline
        \multicolumn{9}{|c|}{{\bf Expected Remaining Lifetime $e(t,a)$}} \\ \hline 
        ~              & \multicolumn{8}{|c|}{{\bf Chronological Age $\kappa_t$}}   \\ \hline
        \textbf{B-Age} & \textbf{60}                                 & \textbf{65}    & \textbf{70}    & \textbf{75}    & \textbf{80}    & \textbf{85}    & \textbf{90}    & \textbf{95}    \\
\hline 
        \textbf{45}             & 31.68                              & 29.18 & 26.60 & 23.93 & 21.17 & 18.29 & 15.27 & 12.09 \\ \hline 
        \textbf{50}             & 29.73                              & 27.46 & 25.12 & 22.68 & 20.14 & 17.48 & 14.67 & 11.68 \\ \hline 
        \textbf{55}             & 27.49                              & 25.49 & 23.41 & 21.23 & 18.94 & 16.52 & 13.95 & 11.19 \\ \hline 
        \textbf{60}             & {\bf 24.95}                              & 23.24 & 21.44 & 19.55 & 17.55 & 15.41 & 13.11 & 10.61 \\ \hline 
        \textbf{65}             & 22.10                              & {\bf 20.70} & 19.22 & 17.64 & 15.95 & 14.12 & 12.13 & 9.92  \\ \hline 
        \textbf{70}             & 18.97                              & 17.90 & {\bf 16.74} & 15.49 & 14.13 & 12.64 & 10.99 & 9.11  \\ \hline 
        \textbf{75}             & 15.67                              & 14.90 & 14.07 & {\bf 13.15} & 12.13 & 10.98 & 9.68  & 8.17  \\ \hline 
        \textbf{80}             & 12.35                              & 11.86 & 11.31 & 10.69 & {\bf 9.99}  & 9.18  & 8.24  & 7.10  \\ \hline 
        \textbf{85}             & 9.22                               & 8.94  & 8.62  & 8.25  & 7.83  & {\bf 7.32}  & 6.70  & 5.93  \\ \hline 
        \textbf{90}             & 6.48                               & 6.35  & 6.19  & 6.01  & 5.79  & 5.51  & {\bf 5.17}  & 4.70  \\ \hline 
        \textbf{95}             & 4.31                               & 4.26  & 4.20  & 4.12  & 4.03  & 3.90  & 3.74  & {\bf 3.51}  \\  
        \hline
    \end{tabular}
\end{center}
\vspace{-0.2in}
\end{table}

Table \ref{erl} provides some estimates of life expectancy under a variety of biological and chronological age assumptions, obtained by solving \eqref{meanlifetimeeqn} numerically. The bolded diagonal values which range from $24.95$ years down to $3.51$ years are comparable to (unisex) life expectancy values from (Canadian) population mortality tables during the ages 60 to 95. 

For example, a 65 year-old retiree who is judged or estimated to be 20 years younger with a biological age of 45, has an expected remaining lifetime of 29.2 years. In contrast, the same 65 year-old with a biological age of 85 (i.e. the retiree is 20 years older than their chronological age) will have a life expectancy of only 8.9 years. Although these values are model dependent, the main assumption being made is that the instantaneous (population) mortality rate of a 60 year-old (whose B and C ages agree) is 0.005 and the instantaneous (population) mortality rate of a 110 year-old is exactly 1, and there is a generalized Brownian bridge that links them.

What this also means (in our model) is that one can in fact grow younger over (short) periods of time. If we think in terms of $[B,C]$ age co-ordinates in the table, a retiree at point $[B=C=65]$ has a life expectancy of $20.70$ years. Ten (calendar) years later their chronological age is 75, but their biological age (might) wander to 60, which represents a substantial improvement in health, a reduction in mortality rate and revised life expectancy of 21.44 years. In a classical actuarial (deterministic) model of aging this would be impossible. Life expectancy would (always) decline over time. 

With stochastic aging this is no longer the case. Life expectancy can (occasionally) increase over time.  Of course the probability that this event will occur and that your biological age will take a favorable path downward depends very much on the underlying parameters. Moreover, computing the survival probabilities requires some additional technology -- a new set of PDEs -- which is addressed in great detail in the technical appendix at the end of this paper.

\section{Consumption over Time and Age \label{LCM}}

With the stochastic mortality model behind us, we are ready to discuss lifecycle economics.

\subsection{Optimal Spending Rate: Deterministic Aging}

We begin with a (very) short review of the canonical lifecycle model under deterministic aging along the lines initiated by Ramsey (1928), Yaari (1964, 1965), Hakansson (1969), Fischer (1973), Richard (1975), Levhari \& Mirman (1977), Davies (1981), Kingston (2000), Butler (2001), Lachance (2012), and many others. See also the work by Bommier (2006) as well as Bommier, Harenberg and Le Grand (2016) in which these sorts of models are criticized on the grounds they imply risk neutrality over random lifetimes.

In all of those models the underlying objective is to maximize discounted utility of consumption over one's remaining lifetime, which can be formally expressed as:
\begin{equation*}
v(t,w)=\max_{c_s}E\left[ \int_{t}^{\infty}e^{-\rho (s-t)}u(c_s)1_{\{s\leq \zeta \}}ds\right],
\end{equation*}
where $\zeta $ is the (random) remaining lifetime and $w$ is household wealth. We set $\Pr [\zeta>s]=p(s,\lambda _{0})$. Without any loss of generality we can assume that $t=0$, since with deterministic mortality rates the problem is essentially time consistent because (unlike what follows in section~\ref{sec:stochasticspending}) no new information about hazard rates will be revealed over time. Moreover, when the mortality hazard rate (i.e. aging) $\lambda_t$ is deterministic, the optimal consumption $c^{\ast}_{s}$ and $1_{\{s\leq \zeta \}}$ are independent, so by Fubini's theorem the objective function can be written as:
\begin{equation*}
v(t,w)=\max_{c_s}\int_{t}^{\infty}e^{-\rho (s-t)}u(c_s)p(s,\lambda _{0})ds.
\end{equation*}

For most of the above-cited literature and all of what follows in this paper, $u(c)=c^{(1-\gamma )}/(1-\gamma)$, which is CRRA utility (although the framework can be generalized to HARA). Our investable universe will not contain mortality credits (or actuarial notes) other than what would be available within an (exogenous) pension income stream $\pi_s$ (which may or may not be constant.) We will in fact restrict to investing risk-free at rate $r$. Let $W_s$ denote wealth at time $s$. The budget constraint is therefore quite straightforward.
\begin{equation*}
\frac{dW_s}{ds}=rW_s+\pi _{s}-c_s.
\end{equation*}
In fact, one doesn't require dynamic programing techniques to solve the problem as it has been formulated. One can use the Calculus of Variations and in particular the Euler-Lagrange (EL) theorem to express the optimal $W_s$ (wealth trajectory), which satisfies a second-order non-homogenous differential equation in regions where $W_s \neq 0$. Although the optimal trajectory of wealth can't be solved explicitly unless $\lambda$ happens to be constant (i.e. no aging), one can express the relevant consumption function quasi-analytically under some continuous mortality rate models. In fact, when the mortality hazard rate $\lambda_s$ is Gompertz (ie $\lambda_s=\lambda_{x+s}^{\text{G}}$, where $x$ is the initial age), as is standard in the actuarial literature, one can obtain an analytic expression for the optimal $c^{\ast}_s$ and $W_s$. See the original work by Leung (2007), as well as Lachance (2012) and Milevsky and Huang (2010) for more on the deterministic mortality model. Here we simply offer some numerical values to set economic intuition. In particular, using an analytic law of mortality helps shed light on the impact of (what we call) {\em longevity risk aversion} $\gamma$ on the optimal consumption and spending strategy. 

Normally the parameter $\gamma$ within these sorts of lifecycle models represents the elasticity of inter-temporal substitution (EIS) or perhaps a general coefficient of relative risk aversion. In our model and formulation it actually captures an attitude towards longevity risk. Specifically, although the probability of survival to any given age (say 100) is fixed, individuals react differently to this number depending on their attitude to longevity risk. Some might worry about a mere 5\% probability of becoming a centenarian and therefore will consume less during the ages 65 to 100 to provision for their old (unlikely) age. Others are more risk tolerant and would be willing to rationally reduce their consumption if-and-when they ever become centenarians. This is distinct from the subjective discount rate $\rho$, which only captures preferences for known consumption today over known consumption tomorrow. Rather, it is a true {\em longevity risk} aversion which isn't necessarily comparable or equal to the attitude or aversion to other exogenous risks. In what follows we will emphasize and show exactly how $\gamma$ affects consumption and wealth depletion rates. In particular when $\pi _{s}=0$ for all $s$ (no pension) the optimal consumption function can be expressed as:
\begin{equation*}
c^{\ast}_s=c^{\ast}_{0}e^{ks}p(s,\lambda _{0})^{1/\gamma },
\end{equation*}
where the new constant $k:=(r-\rho )/\gamma$. The optimal trajectory of wealth is:
\begin{equation*}
W_t =\left( W_0-c^{\ast}_{0}
\int_{0}^{t}e^{(k-r)s}p(s,\lambda_{0})^{1/\gamma }ds\right) 
e^{rt}.
\end{equation*}
And, since (eventually) $W_{\tau}=0$ for some (albeit very large) $\tau$ in a lifecycle model with no bequest motives, the initial consumption and spending rate is therefore:
\begin{equation*}
c^{\ast }_0=\frac{W_0}{\int_{0}^{\tau}e^{(k-r)s}p(s,\lambda_{0})^{1/\gamma } \, ds},
\end{equation*}
under some technical conditions on the relationship between $\rho, r, \gamma$ and mortality parameters. For the sake of basic (deterministic mortality) intuition, and to highlight the role of longevity risk aversion $\gamma$ we now present a few examples to help contrast with our main (later) results. 

\begin{figure}
\caption{{\em Sample path of wealth and consumption for values of $\gamma=1$ (low) and $\gamma=8$ (high), a.k.a. longevity-risk aversion which measures the cautionary provision for old age.}}
\vspace{-0.1in}
\begin{center}
\includegraphics[width=0.45\textwidth]{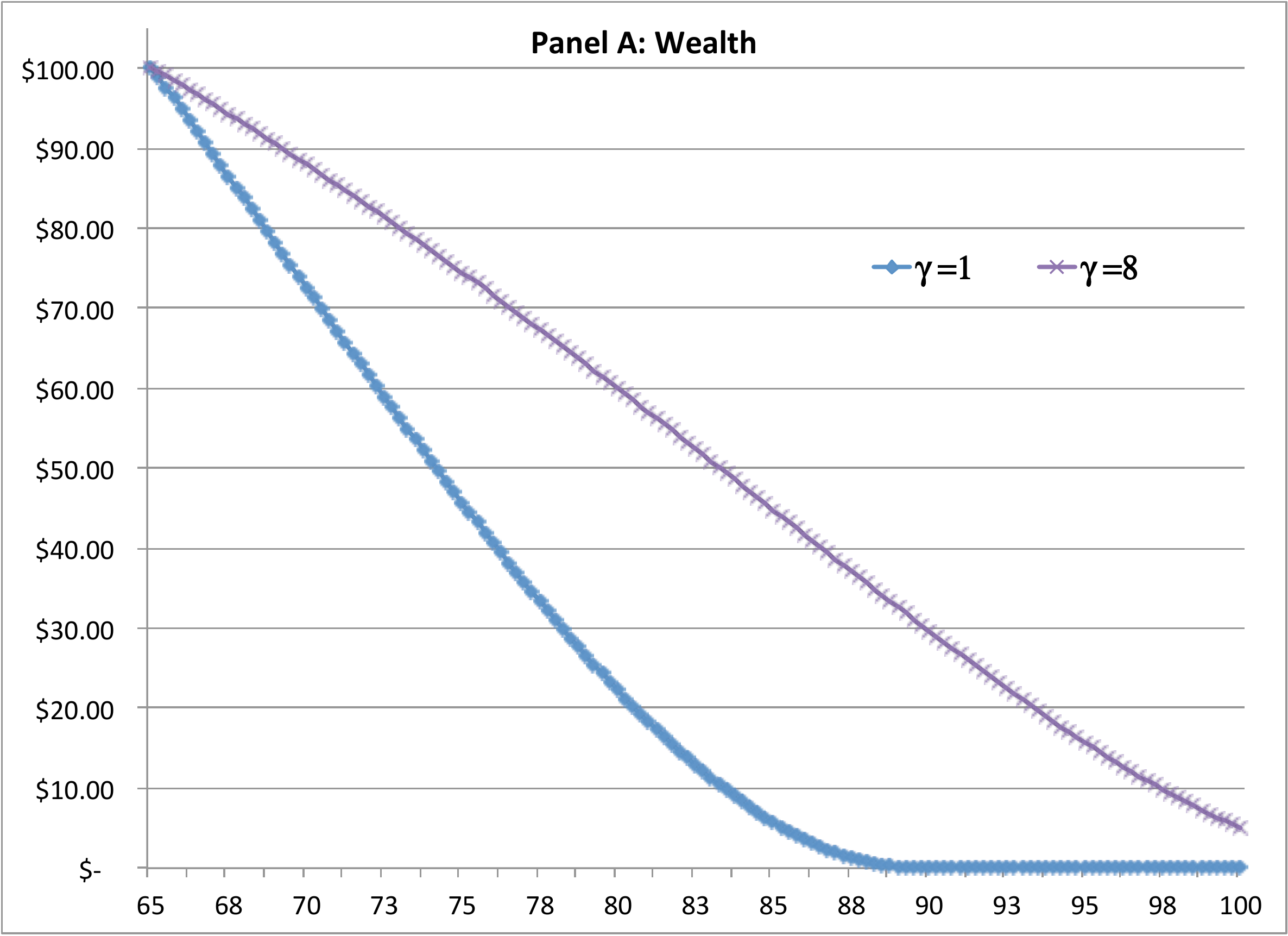} 
\includegraphics[width=0.45\textwidth]{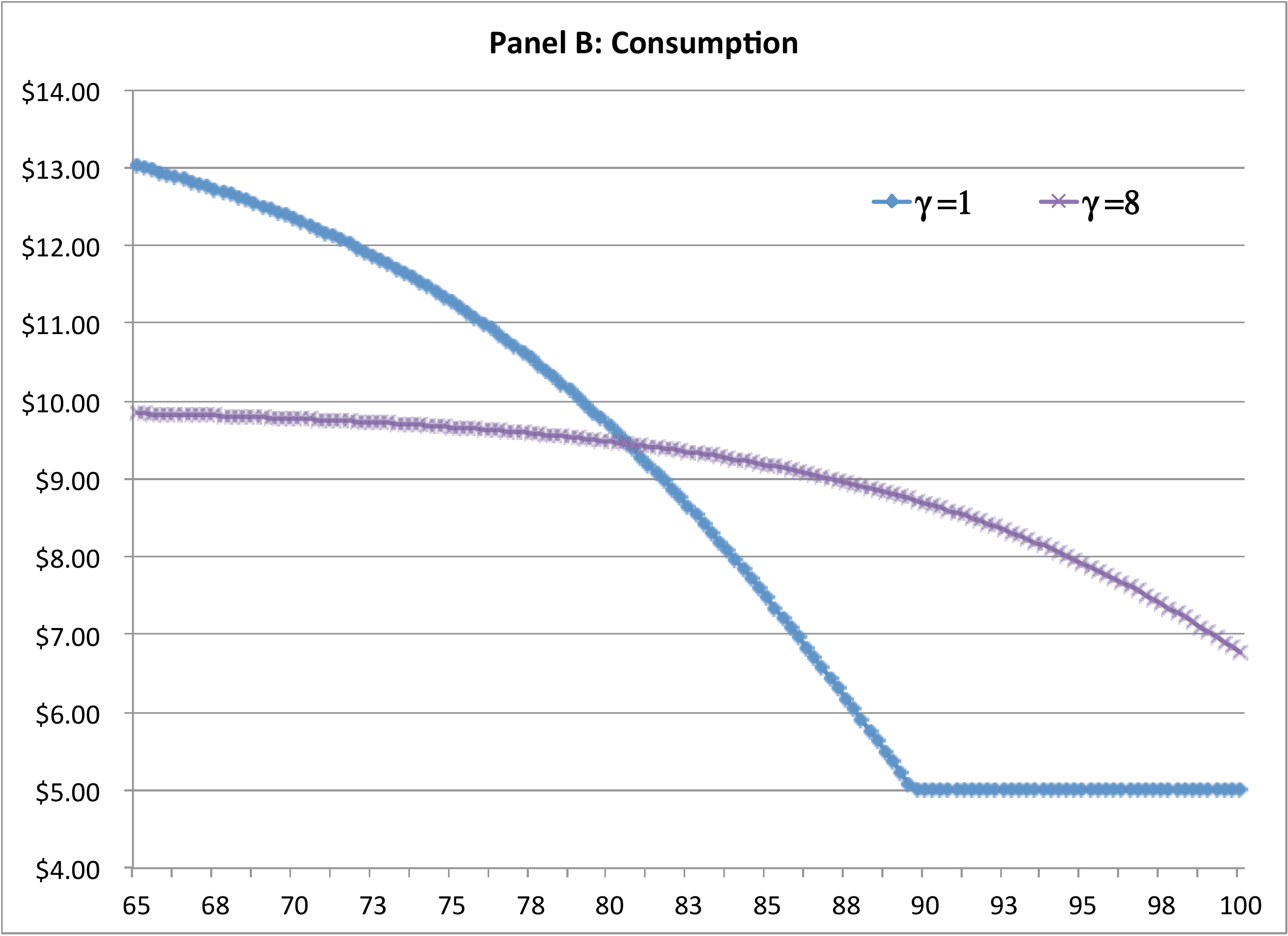} 
\label{CPATH}
\end{center}
\vspace{-0.2in}
\end{figure}

Assume a 65 year-old retiree under a Gompertz law of mortality with mode $m=89.3$ (years) and dispersion: $b=9.5$ (years.) This individual has a current mortality hazard rate of $\lambda_0 = \lambda_{65}^{\text{G}}=\frac{1}{9.5}e^{(65-89.3)/10}=0.00926$, by construction. Under deterministic aging mortality will smoothly increase by $1/(9.5)=10.5\%$ per year to reach a value of $\lambda_{45}=\lambda_{110}^{\text{G}} = \frac{1}{9.5}e^{(110-89.3)/10}=0.83419$ at age of 110, if the retiree is still alive. Our 65 year-old starts retirement with $W_0=\$100$ in consumable wealth and is assumed to have preferences described by CRRA utility with $\gamma=1$ (very low) or $\gamma=8$ (very high) to isolate the impact of this risk aversion. The subjective discount rate $\rho=2.5\%$ in an economy with a real (fixed) interest rate of $2.5\%$ so the two (mostly) cancel each other out. Figure~\ref{CPATH} displays the path of wealth (panel A) and consumption (panel B) over time (during the retirement years) for these two extreme cases. For the purpose of this figure (only) we have added a pension of $\pi_s=\$5$ (for all $s$) to our model -- and solved it numerically -- because it amplifies and highlights the impact of longevity risk aversion $\gamma$ on the qualitative behavior. As mentioned above, an individual with a relatively low level of longevity risk aversion will deplete their wealth much more rapidly, relative to optimizers with high $\gamma$. They will eventually (and rationally) deplete their wealth at a relatively young age and then continue to subsist solely on their pension income. This drawdown or spending plan is set-up at time zero (i.e. retirement age) in the full-knowledge that if-and-when they survive to the wealth depletion time $\tau$ (WDT) their standard of living will drop to the pension level. They are rationally responding to an event with a small probability. 

\subsection{Optimal Spending Rate: Stochastic Aging}
\label{sec:stochasticspending}

So much for the (classic) deterministic mortality and hazard rate model. We now move on to stochastic mortality rates, which allows us to differentiate between Biological and Chronological age. Assume that all wealth $W_t$ is invested at the risk-free rate denote by $r$, thus abstracting from portfolio choice consideration, and with the usual controllable consumption rate $c_t$. Once again utility is assumed to be CRRA($\gamma$) with no utility of bequest and with $\gamma\neq 1$ and a subjective discount rate $\rho$. Once again, we ignore pension income (which would have required us to compute a wealth depletion time) and assume that all consumption comes from portfolio withdrawals.

Let $v(t,a,w)$ be the value function for maximizing utility of consumption, which includes both time $t$ and Biological age $a$. Namely; 
\begin{equation}
v(t,a,w)=\sup_{c_s}E\Big[\int_t^{\infty}e^{-\rho (s-t)} \frac{c_s^{1-\gamma}}{1-\gamma}1_{\{s<\zeta\}}\,ds\mid A_t=a, W_t=w\Big],
\label{objective2}
\end{equation}
with the same simple budget constraint we introduced in the prior section.
\begin{equation*}
dW_t=(rW_t-c_t)\,dt
\end{equation*}

The derivation now is trickier and we cannot use Calculus of Variation arguments due to the stochasticity of the mortality rate. We must resort to martingale methods and refer the interested reader to the derivations and proofs in the Appendix. For now, we note that $v(t,a,w)= f(t,a)w^{1-\gamma}/(1-\gamma)$ and the optimal consumption is $c=w f(t,a)^{-\frac{1}{\gamma}}$. This will obviously be familiar to economists because the marginal utility $u^{\prime}(c) = v_{w}(t,a,w)$ in the basic lifecycle framework.

We now move on to state a partial differential equation (PDE) for the optimal consumption function, which can be expressed in the following way.
\begin{equation}
f_t+\Big(1+\xi\frac{\kappa_t-a}{T-t}\Big)f_a + \frac{\sigma^2}{2}f_{aa}+r(1-\gamma)f-(\rho+\lambda(a))f+\gamma f^{1-\frac{1}{\gamma}}=0
\label{consumptioneqnfform}
\end{equation}
for $t<T$, with boundary conditions $f(T,a)=f_T$, $f(t,\infty)=0$. For the boundary condition at $a=-\infty$ we have the simpler problem where $f$ does not depend on $a$, and $\lambda=0$. In other words, \eqref{consumptioneqnfform} becomes an ODE 
\begin{equation*}
f_t+r(1-\gamma)f-\rho f+\gamma f^{1-\frac1\gamma}=0.
\end{equation*}
Note that for $t\ge T$, $f$ does not depend on $a$. Therefore the dynamics are time invariant, which means $f$ is a constant $f_T$. It is determined by the following equation:
\begin{equation*}
r(1-\gamma)f_T-(\rho+\lambda_T)f_T+\gamma f^{1-\frac{1}{\gamma}}_T=0,
\end{equation*}
which is obtained by setting $f_t=f_a=f_{aa}=0$ in equation (\ref{consumptioneqnfform}). From this we can obtain the following explicit solution 
\begin{equation*}
f_T=\Big(\frac{\rho+\lambda_T-r(1-\gamma)}{\gamma}\Big)^{-\gamma}.
\end{equation*}
Now \eqref{consumptioneqnfform} can be solved numerically. This will be one ingredient in the results to be described in section \ref{NUM}.

The above applies with $\lambda\neq 1$. There are related equations that give $v$ in the logarithmic case $\lambda=1$, for which we refer to \eqref{consumptioneqnfformlogarithmic} and \eqref{consumptioneqnhformlogarithmic} of the Appendix. 

Finally there is a stability condition that is needed when $0<\gamma<1$, namely that the interest rate $r$ can't be much larger than the subjective discount rate $\rho$, which is actually a standard restriction imposed in similar lifecycle models. In particular note that if $r\gg \rho$, the rational consumer optimally defers consumption forever. For most of the numerical examples that follow (to flush out the impact of mortality) we will assume that $\rho=r$, so these stability issues will not be a concern.

\subsection{Comparative Statics}

Equation (\ref{consumptioneqnfform}) contains all the information we need to establish the optimal spending (or consumption) rate at any given [C,B] age pair. So, before we move on to numerical examples and the empirical calibration of our model, in this subsection we discuss (what we know about) the impact of the mortality parameters $\lambda_0, \lambda_T$, diffusion parameters $\xi, \sigma$, preference parameters $\rho,\gamma$ and the interest rate $r$ on the optimal consumption rate. How do these seven parameters affect optimal behavior? 

As far as subjective discount rate $\rho$ and relative risk aversion (or inter-temporal elasticity of substitution) $\gamma$ is concerned, our stochastic mortality model is consistent with standard results. A higher value of $\rho$ (a.k.a. impatience) will increase current consumption at the expense of future consumption, no different from a deterministic Yaari (1965) model and its many extensions. With regards to $\gamma$, a higher value will lead to more cautious behavior -- or greater longevity risk aversion -- so current consumption rates should be reduced. Numerical results support this. An increase in the interest rate $r$ when $\gamma > 1$ will increase the initial (current) consumption rate. The opposite occurs when $\gamma < 1$. All standard in the economics literature.

As far as the biology is concerned, raising the mortality rate curve lowers the probability of surviving to any given age, which in turn increases the optimal consumption rate for a rational consumer. In terms of our model parameters, increasing either or both of the initial or terminal mortality rates $\lambda_0$ or $\lambda_T$ raises the mortality curve except at its extremities, so it should still increase spending, as numerical results indeed support. 

Again, the impact of a shift in {\em deterministic} mortality rates is well-known in the standard lifecycle literature and was originally studied and emphasized by Levhari and Mirman (1977).  With regards to the impact of volatility of mortality $\sigma$ and the speed of mean reversion $\xi$ -- which are the two parameters driving the diffusion for the biological age $A_t$ in equation (\ref{dynamicsforA}) -- the situation is more complex. Both high $\sigma$ and low $\xi$ should have similar effects, in that they lead to more wandering for biological age. What we see in numerical experiments is that, for a fixed value of C-age, high $\sigma$ will increase optimal consumption a small amount when B-age is low, but will lower consumption (by a somewhat larger margin) when B-age is high. The sensitivity to $\sigma$ is very small when B-age equals C-age, but the crossover does not appear to precisely coincide with having the ages match. 

An indication that the qualitative behaviour can be more subtle comes from Huang, Milevsky and Salisbury (2012). That paper suggests that if we were to raise $\sigma$ while holding unconditioned survival probabilities fixed, then consumption rates would rise when $\gamma>1$ but decline when $\gamma<1$ (less risk-averse individuals may opt to stay invested, relying on their ability to ramp consumption up later, if their B-age grows rapidly). 

A deeper analysis of the qualitative dependence on $\sigma$ and $\xi$ requires further investigation. It does seem to be the case that the sensitivity of consumption to changes in B-age (holding $\sigma$ fixed) is more significant than its sensitivity to the volatility $\sigma$ of B-age (holding B-age fixed). 

\subsection{Dispersion in Future Spending Rates}

Recall that in addition to the {\em normative} objectives of this paper, we also wanted to show that {\em even} in a relatively simple life-cycle model with {\em no} bequest motives, {\em no} portfolio choice decisions and identical risk preference parameters $\rho$ and $\gamma$, one might still observe a dispersion in retirement spending rates for a homogenous group of individuals who are at the same chronological age today. This dispersion would arise solely because the retiree's biological age -- which is necessary but not sufficient for consumption decisions -- will itself vary or wander over time.  Recall that by time $t=T$ the wandering will cease and the spending rate (assuming our retiree is still alive) will converge to the known $f(T,\kappa_T=A_T)$, because the aging process stops. 

Our plan in this section is to describe the methodology for computing the magnitude of this dispersion over time (and then offer some numerical examples). To think about this in practical terms, we start our canonical retiree with a chronological and biological age of $\kappa_0=60$ and corresponding mortality rate of $\lambda_0=0.005$. We know that if-and-when he/she reaches the chronological age of $\kappa_{50}=110$ the mortality rate will plateau at $\lambda_{50}=1$ from there onward; these parameters are the basis for most of our numerical examples.

At time $t=5, t=10, t=20, t=30$ the chronological age will (obviously) be $\kappa_5 =65, \kappa_{10} =70, \kappa_{20} =80, \kappa_{30}=90$ respectively, but the biological age ($a$) is a random variable with a (sub) probability density function denoted by $g(t,a).$ In some sense, that density is the conceptual core on which this paper is based. 

The (sub-density) function $g(t,a)$ will satisfy what is known as a {\em forward equation} and the appendix contains a derivation of the PDE satisfied by $g(t,a)$, which can be written as:
\begin{equation}
g_t(t,a)+\frac{\partial}{\partial a}\Big(\Big(1+\xi\frac{\kappa_t-a}{T-t}\Big)g(t,a)\Big)
-\frac{\sigma^2}{2}g_{aa}(t,a)+\lambda(a)g(t,a) = 0.
\label{subg}
\end{equation}
This equation, like the PDE for the optimal consumption and spending rate, can be solved using numerical techniques.

Now, let $\alpha(t,q)$ denote the $q$'th percentile of the biological age at time t conditional on being alive, that is $\Pr[A_t \leq\alpha(t,q) | \, survival ]=q$. For example the expression $\alpha(15,0.95)=75$ means that if you live to chronological age 75, there will be a 95\% (conditional) probability that your biological age will be less than 75 at that time; assuming that your current biological and chronological age is 60. Likewise, the range from 
$\alpha(15,0.05)$ to $\alpha(15,0.95)$ would provide a 90\% confidence interval for the biological age of a survivor at time $t=15$, etc., which is what we (tried to) plot in Figure~\ref{BvC}

So much for how $\alpha(t,q)$ is defined and what it represents, now let's discuss how it is actually computed. By the formal definition of $q$ as a probability and the properties of the probability sub-density function $g(t,a)$, we know that: 
\begin{equation}
q = \frac{1}{S(t)} \int_{-\infty}^{\alpha(t,q)} g(t,a) \, da
\label{eq.q}
\end{equation}
where $S(t)=\int_{-\infty}^{\infty} g(t,a) \, da$ is a scaling factor.  Then, differentiating both sides of equation~(\ref{eq.q}) with respect to $q$ leads to the equality: $1 = \frac{1}{S(t)} g(t,\alpha(t,q)) \frac{\partial \alpha(t,q)}{\partial q}$. and the partial derivative term (that is the probability density function for biological age conditional on survival, can be computed as:
$ \frac{\partial \alpha(t,q)}{\partial q} = S(t)/g(t, \alpha(t,q)). $

Finally, since spending varies monotonically with age, substituting $\alpha(t,q)$ into the optimal spending rate function, namely $f(t,\alpha(t,q))^{-1/\gamma}$, will provide us with the $q$'th percentile for spending at time $t$. The range $f(t,\alpha(t,0.95))^{-1/\gamma}$ to $f(t,\alpha(t,0.05))^{-1/\gamma}$ is the 90\% confidence interval for the dispersion of spending rates at time $t$ (chronological age $\kappa_t$), conditional on survival. 

In other words the above enables us to map or convert the uncertainty or dispersion in observed biological ages -- at any given and fixed chronological age -- to a range of plausible value for withdrawal rates and spending rates.

\newpage

\section{Testable Implications \label{NUM}}

\subsection{Spending rates}

Table~\ref{spending1} and Table~\ref{spending2} provide an assortment of numerical values for spending rates under a variety of chronological and biological ages. For example let's start with a (chronological) $\kappa_0=60$ year-old whose biological age is (also) measured at $A_0=60.$ Assume that we are in an $r=2.5\%$ interest rate environment which is identical to this individual's subjective discount rate $\rho=2.5\%$. The relevant mortality rate parameters are $\lambda_0=0.005$ and $\lambda_{110}=1.0$. 

Here is how to interpret the results: A consumer with a very high level of longevity risk aversion (i.e. $\gamma=8$) will consume or spend at a rate of $3.834\%$ of wealth under these parameter conditions. In contrast, but under the same parameter conditions, a consumer with a relatively modest level of longevity risk aversion (i.e. $\gamma=2$) will spend at a higher rate of $4.798\%$. There are no surprises so far and this is exactly what one might expect in the classical (deterministic aging) model which we described earlier in Section~\ref{LCM}.

\begin{table}
\caption{{\em The optimal retirement spending rate (computed via $f^{-1/\gamma}$) as a function of biological and chronological age assuming $\gamma=8$ and $\rho=r=2.5\%$. The underlying mortality rate is $\lambda=0.005$ at age 60 and pinned at $\lambda=1.0$ at age 110, with $\xi=1$ and $\sigma=30\%$.}} 
\label{spending1}
\begin{center}
    \begin{tabular}{|c|c|c|c|c|c|c|c|c|}
    \hline
      \multicolumn{9}{|c|}{{\bf Optimal Spending Rates: HIGH Longevity Risk Aversion}} \\ \hline 
            ~              & \multicolumn{8}{|c|}{{\bf Chronological Age $\kappa$}}   \\ \hline
        \textbf{B-Age} & \textbf{60}                                 & \textbf{65}    & \textbf{70}    & \textbf{75}    & \textbf{80}    & \textbf{85}    & \textbf{90}    & \textbf{95}    \\ \hline 
\hline 
        \textbf{45}             & 3.638\%                              & 3.800\% & 3.998\% & 4.246\% & 4.563\% & 4.980\% & 5.551\% & 6.374\% \\ \hline 
        \textbf{50}             & 3.688\%                              & 3.852\% & 4.051\% & 4.301\% & 4.620\% & 5.038\% & 5.611\% & 6.435\% \\ \hline 
        \textbf{55}             & 3.752\%                              & 3.917\% & 4.118\% & 4.369\% & 4.690\% & 5.110\% & 5.684\% & 6.509\% \\ \hline 
        \textbf{60}             & 3.834\%                              & 4.000\% & 4.203\% & 4.456\% & 4.778\% & 5.200\% & 5.775\% & 6.601\% \\ \hline 
        \textbf{65}             & 3.942\%                              & 4.110\% & 4.314\% & 4.568\% & 4.892\% & 5.316\% & 5.892\% & 6.717\%  \\ \hline 
        \textbf{70}             & 4.087\%                              & 4.256\% & 4.461\% & 4.717\% & 5.042\% & 5.466\% & 6.042\% & 6.865\%  \\ \hline 
        \textbf{75}             & 4.285\%                              & 4.455\% & 4.662\% & 4.918\% & 5.243\% & 5.667\% & 6.242\%  & 7.061\%  \\ \hline 
        \textbf{80}             & 4.565\%                              & 4.735\% & 4.941\% & 5.197\% & 5.521\%  & 5.943\%  & 6.513\%  & 7.323\%  \\ \hline 
        \textbf{85}             & 4.969\%                               & 5.137\%  & 5.342\%  & 5.595\%  & 5.915\%  & 6.330\%  & 6.891\%  & 7.684\%  \\ \hline 
        \textbf{90}             & 5.571\%                               & 5.735\%  & 5.934\%  & 6.180\%  & 6.490\%  & 6.892\%  & 7.433\%  & 8.195\%  \\ \hline 
        \textbf{95}             & 6.502\%                               & 6.655\%  & 6.841\%  & 7.070\%  & 7.359\%  & 7.735\%  & 8.239\%  & 8.945\%  \\
        \hline
    \end{tabular}
\end{center}
\end{table}

Let's now front-forward by 10 years when our consumer (retiree) is aged 70 chronologically.   Within our lifecycle model their optimal spending rate depends (also) on biological age since chronological age is no longer enough information (a.k.a. a sufficient statistic.)  If their age co-ordinates happen to be: [C=70, B=70] then their spending rate is $4.461\%$ in Table~\ref{spending1}, which is the case of high risk aversion. But if they have ``aged well'' and their biological age is only $a=65$, their age co-ordinates in the matrix are now [C=70,B=65] and the optimal spending rate is (a relatively lower) $4.314\%$. The 15 basis point difference between the {\em average} 70 year-old and the {\em young} 70-year old might not seem like much, but recall that these values depend on subjective preference parameters $\gamma,\rho$, as well as mortality parameters $\lambda_0, \lambda_T$ and interest rates $r$. The difference between old and young could be (much) greater depending on specific values. For example, and in the same table, at the chronological age of 80 the lowest spending rate is $5.56\%$ and the highest spending rate is more than triple at $16.25\%$ The odds of biological age wandering to these levels are slim but the point remains.

\begin{table}
\caption{{\em Same parameters as table \ref{spending1}, but with $\gamma=2$.}} 
\label{spending2}
\vspace{-0.25in}
\begin{center}
    \begin{tabular}{|c|c|c|c|c|c|c|c|c|}
    \hline
        \multicolumn{9}{|c|}{{\bf Optimal Spending Rates: LOW Longevity Risk Aversion}} \\ \hline 
            ~              & \multicolumn{8}{|c|}{{\bf Chronological Age $\kappa$}}   \\ \hline
        \textbf{B-Age} & \textbf{60}                                 & \textbf{65}    & \textbf{70}    & \textbf{75}    & \textbf{80}    & \textbf{85}    & \textbf{90}    & \textbf{95}    \\
\hline 
        \textbf{45}             & 4.242\%                              & 4.465\% & 4.743\% & 5.096\% & 5.560\% & 6.195\% & 7.116\% & 8.572\% \\ \hline
        \textbf{50}             & 4.379\%                              & 4.607\% & 4.889\% & 5.247\% & 5.717\% & 6.359\% & 7.289\% & 8.757\% \\ \hline
        \textbf{55}             & 4.559\%                              & 4.791\% & 5.078\% & 5.442\% & 5.918\% & 6.569\% & 7.509\% & 8.989\% \\ \hline
        \textbf{60}             & 4.798\%                              & 5.035\% & 5.328\% & 5.698\% & 6.182\% & 6.841\% & 7.792\% & 9.286\% \\ \hline
        \textbf{65}             & 5.123\%                              & 5.366\% & 5.665\% & 6.043\% & 6.535\% & 7.203\% & 8.166\% & 9.675\%  \\ \hline
        \textbf{70}             & 5.579\%                              & 5.827\% & 6.133\% & 6.518\% & 7.019\% & 7.697\% & 8.672\% & 10.196\%  \\ \hline
        \textbf{75}             & 6.233\%                              & 6.487\% & 6.800\% & 7.192\% & 7.701\% & 8.389\% & 9.374\%  & 10.912\%  \\ \hline
        \textbf{80}             & 7.203\%                              & 7.462\% & 7.779\% & 8.177\% & 8.691\%  & 9.386\%  & 10.379\%  & 11.923\%  \\ \hline
        \textbf{85}             & 8.691\%                               & 8.951\%  & 9.268\%  & 9.666\%  & 10.180\%  & 10.873\%  & 11.863\%  & 13.400\%  \\ \hline
        \textbf{90}             & 11.054\%                               & 11.306\%  & 11.614\%  & 12.000\%  & 12.500\%  & 13.175\%  & 14.138\%  & 15.635\%  \\ \hline
        \textbf{95}             & 14.937\%                               & 15.166\%  & 15.447\%  & 15.800\%  & 16.258\%  & 16.878\%  & 17.768\%  & 19.159\%  \\
        \hline
    \end{tabular}
\end{center}
\vspace{-0.2in}
\end{table}

\begin{figure}
\caption{{\em Spending Surface: Optimal lifecycle spending as a function of two ages.}}
\vspace{-1.5in}
\begin{center}
\includegraphics[width=0.7\textwidth]{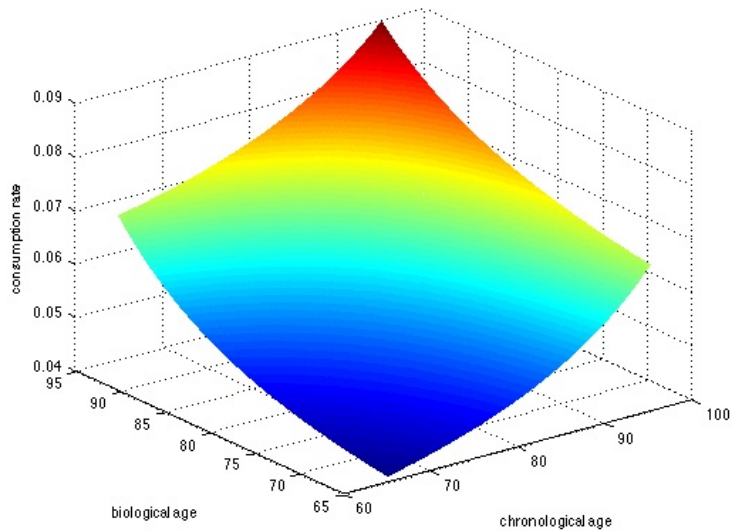}  
\end{center}
\label{SURFACE}
\vspace{-1.75in}
\end{figure}

Figure \ref{SURFACE} offers another perspective on the impact and importance of both age co-ordinates on the optimal spending rate function. The horizontal plane represents the biological and chronological ages and the vertical dimension is the corresponding spending rate for those co-ordinates. Notice that in the near corner where the individual is relatively young the spending rate as a function of wealth is in the $4\%$ range. As the individual ages in either chronological (to the right) or biological (to the left), the spending rate as a function of wealth increase. However one can see the gradient or slope is higher as one ages biologically as opposed to chronologically, all else being equal of course.

Finally, as we reach the summit or upper back corner of the graph when chronological age is 100 and biological age is 95, the spending rate function $f^{-1/\gamma}$ reaches a peak of approximately $9\%$, which corresponds to the numbers at the lower right corner of Table~\ref{spending1} or Table~\ref{spending2}. Once again we caution the reader that these numbers and figures are quite sensitive to parameter values and therefore one shouldn't read too much into the fact they are within the vicinity of the so-called $4\%$ rule that has been widely advocated by the media and financial advisors. In sum, if indeed such a model were to be implemented in a normative context -- and to provide financial advice -- the user would have to take great care in estimating appropriate consumer specific values of (longevity) risk aversion $\gamma$, subjective discount rates $\rho$ and a suitable long-term real-return $r.$ 

\subsection{Comparing models}

How might we calibrate the parameters of the model for Biological age and test its goodness of fit? Indeed, testing should become more feasible in the near future since current information about biological age is limited and subjective which also limits the ability to control consumption on the basis of biological age. That will change as high-quality measurements of biological age and testing become widespread. Longitudinal data on individuals might eventually become the gold standard -- either directly measuring how biological age evolves in individuals, or measuring this indirectly using economic data on the evolution of consumption.

At present it seems more realistic to explore the potential for testing our models based on statistical evidence about population spending rates. For example, consider two alternative explanations for the dispersion in spending rates at a given chronological age. One is that biological age evolves stochastically over time, as we have modeled it ({\it stochastic ageing}). An alternative model ({\it deterministic adult ageing}) might imagine a deterministic evolution of biological age (or equivalently of hazard rates), but starting from a random value. For example, could it be that dispersion in biological ages is due entirely to influences early in life, so that two individuals whose biological ages match at chronological age 50 (say) will also match at subsequent ages?

We don't yet have data with which to make that comparison. But we can illustrate how it might be carried out. Fix values of $r$, $\rho$, $\gamma$, $\xi$, $a_T$, $m$ and $b$, leaving only $\sigma$ to vary. To compare a stochastic ageing model (eg. with $\sigma=0.3$) and a deterministic ageing model (with $\sigma=0$) we require two confidence intervals for consumption (e.g. one at C-age 60 and another at C-age 85). Using the solutions to equation (\eqref{consumptioneqnfform}), we translate the first of these into a confidence interval for B-age in each model. In each model we may fit to this interval using an appropriate choice for the initial condition in equation (\eqref{subg}). Now solve \eqref{subg} forward to determine a B-age confidence interval at the later time and from that the corresponding confidence interval for consumption. We should now be able to see which model fits the data better. 

Another way of saying this is that two confidence intervals for observed consumption rates should allow us to calibrate $\sigma$ (assuming given values for the other parameters). In principle, the same approach should let us calibrate multiple parameters, or compare different models for stochastic ageing. For example, to calibrate $\sigma$ and $\xi$ simultaneously might require three such intervals (e.g. for consumption at ages 60, 75, and 85), although the relatively low sensitivity of consumption to $\sigma$ might make it more difficult to obtain robust estimates from consumption itself. Biological data and metrics would obviously be preferable. 

\section{Conclusion \label{CON}}

In this paper we have taken the first step towards constructing a lifecycle utility maximization model -- which is the {\em working horse} in the retirement literature -- in which the agent's age does not move in lockstep with calendar time. Our main break with the past economic literature is that we assume a world in which (true) biological age can (i.) be measured with precision and (ii.) grows at a stochastic non-linear rate which might even decline over periods of time. In our initial (introductory) model we assume Biological age is exogenous and can not be influenced, modified or controlled. To put it bluntly, you can't do anything to improve or harm your health.

Moreover, in the future Biological age might be measured with a single biomarker of aging, such as the telomeres we mentioned in the introduction or perhaps using a much more comprehensive inventory of {\em deficits}, such as those documented by Mitinski, Bao and Rockwood (2006). Our point isn't to advocate a particular methodology for measuring biological age, but rather to (i.) introduce economists to the fact that it is distinct from chronological age, (ii.) discuss its unique stochastic properties, and finally (iii.) offer a plausible stochastic process that can be used in a reformulated {\em economic} lifecycle model.  We focused exclusively on the retirement stage (i.e. when human capital is exhausted) to isolate and examine the impact of aging, which is the period when mortality rates have a noticeable impact on optimal behavior. Indeed, longevity uncertainty would be expected to drive consumption rates, much more so than during the working years. 

We re-iterate that at some level our work is (also) about how relative health status affects retirement spending and consumption rates. However in this paper we {\bf frame} the problem very differently. We position everyone as having a clearly identifiable and known biological age, a number that might soon become easily available with technological advances, wearables, etc. As a link to the prior literature, we modify the lifecycle optimality conditions and show how to derive the proper consumption, withdrawal or spending rate function in a Yaari (1965) model with stochastic aging. This introduces another state variable into the (partial differential) equation, namely ones biological age. If indeed retirees soon have access to (a device with) their biological age in real time then our model would inform them how to draw-down (spend, consume) their assets in a way that is consistent with a rational lifecycle model. As we showed, they would require both their biological and chronological age to make these economic decisions.

We believe that in addition to a contribution to {\em normative} theory, our model helps explain the cross-sectional heterogeneity in retirement spending rates at any given chronological age -- a phenomenon which has been documented in various countries -- by suggesting that this dispersion is due to (perceived) differences in biological age. For example, under one set of parameters we show that the optimal spending rate can vary in the range of $5\%$ to $15\%$ at retirement -- for a fixed value of longevity risk aversion ($\gamma$) and no bequest motives -- when real long term (risk free) investment rates are $\rho=r=$2.5\%. The exact number is sensitive to the usual financial and economic inputs and is not really the point of our analysis. 

\subsection{Next Steps}

This is the beginning of a research agenda to investigate the implications of {\em acting your age} in a rational lifecycle framework. Future research might involve introducing risky assets to the choice set as well as (properly) priced life annuities. This would allow us to investigate the impact of B-age vs. C-age on portfolio choice, the demand for longevity insurance as well as optimal consumption. In addition to the normative implications, such an extension would also help shed light on the so-called annuity puzzle, which is the low demand for (voluntary) annuities. This is an area of fertile research, see for example Feigenbaum, Gahramanov and Tang (2013) for a paper discussing the various aspects and dimensions of this puzzle.  Likewise, by introducing a risky asset -- such as in Kingston (2000) who extends the Merton model -- we could investigate the impact of flexible vs. fixed retirement dates or the choice between labour vs. leisure.

We conclude by musing whether on a societal level it might make sense to subordinate retirement policy itself to biological age as well as chronological age. Note that this could be varied across retirement schemes, leaving compulsory national pensions (where true risk-pooling is possible) based solely on chronological age, but allowing employment-based pensions to use both ages.  For example, as an alternative to extending or increasing pension eligibility ages -- which has its own political problems and fears of redistribution -- perhaps people should be allowed to draw pensions (only) at the biological age of 65. 

{\em Might B-age retirement policies be perceived as more equitable than those based on C-age?} We leave for future research...

\pagebreak

\pagebreak

\section{Appendix: Derivations and Proofs}

\subsection{Derivation of the PDE Satisfied by Optimal Consumption}

Our starting point here is the objective function in equation \eqref{objective2}. We start by assuming there is an optimal choice of the control $c_s$. For that choice, consider 
\begin{equation*}
E[\int_0^\zeta e^{-\rho s} \frac{c_s^{1-\gamma}}{1-\gamma}\,ds \mid \mathcal{F}_t].
\end{equation*}
The portion of the integral before $t$ is $\mathcal{F}_t$-measurable so can be moved outside the expectation, and what remains is a modification of equation \eqref{objective2}.
In other words, for $t<\zeta$, and the optimal $c_t$, we have that:
\begin{equation}
\int_0^t e^{-\rho s} \frac{c_s^{1-\gamma}}{1-\gamma}\,ds + e^{-\rho t}v(t,A_t,W_t)
=E[\int_0^\zeta e^{-\rho s} \frac{c_s^{1-\gamma}}{1-\gamma}\,ds \mid \mathcal{F}_t].
\label{martsubmartexpression}
\end{equation}
The RHS of equation \eqref{martsubmartexpression} is a martingale and has a jump of $-e^{-\rho t}v(t,A_t,W_t)$ at $t=\zeta$. The difference between this jump and its compensator is also a martingale, so subtracting it from the RHS of equation \eqref{martsubmartexpression} gives a continuous martingale (see section \ref{sec:means} for a similar argument). Therefore
\begin{equation*}
\int_0^t e^{-\rho s} \frac{c_s^{1-\gamma}}{1-\gamma}\,ds + e^{-\rho t}v(t,A_t,W_t)-\int_0^t \lambda_s e^{-\rho s}v(s,A_s,W_s)\,ds
\end{equation*}
is a martingale when stopped at $\zeta$. Applying It\^o's lemma now shows that for $t<T$,
\begin{equation}
\frac{c_t^{1-\gamma}}{1-\gamma}+v_t+\Big(1+\xi\frac{\kappa_t-a}{T-t}\Big)v_a + \frac{\sigma^2}{2}v_{aa}+(rw-c_t)v_w-(\rho+\lambda(a))v=0
\label{martinoptimalcase}
\end{equation}
in the optimal case. For a sub-optimal choice of $c_t$, a similar argument shows that we get a supermartingale instead, so the LHS of equation \eqref{martinoptimalcase} is $\le 0$. The Hamilton-Jacobi-Bellman (HJB) equation is therefore that:
\begin{equation*}
\sup_c\frac{c^{1-\gamma}}{1-\gamma}+v_t+\Big(1+\xi\frac{\kappa_t-a}{T-t}\Big)v_a + \frac{\sigma^2}{2}v_{aa}+(rw-c)v_w-(\rho+\lambda(a))v=0
\end{equation*}
Therefore $c^{-\gamma}-v_w=0$ and as usual, this becomes
\begin{equation*}
v_t+\Big(1+\xi\frac{\kappa_t-a}{T-t}\Big)v_a + \frac{\sigma^2}{2}v_{aa}+rwv_w+\frac{\gamma}{1-\gamma}v_w^{1-\frac{1}{\gamma}}-(\rho+\lambda(a))v=0.
\end{equation*}
The natural scaling relation is that $v(t,a,kw)=k^{1-\gamma}v(t,a,w)$. 
This implies that 
\begin{equation}
v(t,a,w)=f(t,a)\frac{w^{1-\gamma}}{1-\gamma}
\label{scaledequation}
\end{equation} 
for some $f$, from which we get
\begin{equation*}
f_t+\Big(1+\xi\frac{\kappa_t-a}{T-t}\Big)f_a + \frac{\sigma^2}{2}f_{aa}+r(1-\gamma)f-(\rho+\lambda(a))f+\gamma f^{1-\frac{1}{\gamma}}=0
\end{equation*}
for $t<T$, with boundary conditions $f(T,a)=f_T$, $f(t,\infty)=0$. In other words, we have obtained \eqref{consumptioneqnfform}.
 Optimal consumption then is $c=w\cdot f(t,a)^{-\frac{1}{\gamma}}$. 
 
To remove the assumption that there is an optimal control, we invoke a standard verification theorem argument, that starts with existence of a solution to equation \eqref{consumptioneqnfform}, and concludes that $v$ given by equation \eqref{scaledequation} solves the original optimization problem, and that the $c$ obtained above is in fact optimal. 

\subsection{Special Case of Logarithmic Utility \label{LOGU}}

In the special case of logarithmic utility, which is the limiting case $\gamma\to 1$ for CRRA utility, the same arguments used in the body of the paper (section ~\ref{LCM}) lead to:
\begin{equation}
v_t+\Big(1+\xi\frac{\kappa_t-a}{T-t}\Big)v_a + \frac{\sigma^2}{2}v_{aa}+rwv_w-1-\log v_w-(\rho+\lambda(a))v=0.
\label{consumptioneqnvformlogarithmic}
\end{equation}
In this case, the natural scaling relation is that $v(t,a,kw)=v(t,a,w)+(\log k)E[\int_0^{\zeta-t}e^{-\rho s}\,ds\mid\mathcal{F}_t]$, which implies a solution of the form $v(t,a,w)=f(t,a)\log w + h(t,a)$. Substituting into \eqref{consumptioneqnvformlogarithmic} we obtain
\begin{align}
f_t+\Big(1+\xi\frac{\kappa_t-a}{T-t}\Big)f_a + \frac{\sigma^2}{2}f_{aa}+1-(\rho+\lambda(a))f&=0
\label{consumptioneqnfformlogarithmic}\\
h_t+\Big(1+\xi\frac{\kappa_t-a}{T-t}\Big)h_a + \frac{\sigma^2}{2}h_{aa}-(\rho+\lambda(a))h+rf-\log f-1&=0
\label{consumptioneqnhformlogarithmic}
\end{align}
for $t<T$. Boundary conditions are $f(T,a)=f_T$ and $h(T,a)=h_T$. For $t\ge T$, $f$ and $h$ are constant, so $f_T=\frac{1}{\rho + \lambda_T}$ and $h_T=\frac{1}{\rho + \lambda_T}(rf_T-\log f_T-1)$. Optimal consumption is $c=w\cdot f(t,a)$. 

Finally, in this logarithmic $\gamma=1$ case, the ODE for $f(t)$, $t\ge T$ is $f_t+1-\rho f=0$, which has solution $f(t)=e^{-\rho(T-t)}(f(T)-\frac{1}{\rho})+\frac{1}{\rho}$. There is a similar ODE for $h(t)$, namely 
$h_t-\rho h+rf-\log f-1=0$, which one could solve to obtain the maximized value of utility, necessary for welfare calculations.

\subsection{Approximate Analytic Solution \label{APPROX}}

Consider the main HJB equation \eqref{consumptioneqnfform} for consumption or the spending rate, after a dimension reduction. Note that the diffusion effect within the PDE occurs at O($\sigma^2$), which is smaller in magnitude than the effect in the drift, especially early on when $(T-t)$ is small. Therefore we can approximate the solution to the PDE by taking the following asymptotic expansion:
\begin{equation*}
f\sim f^{(0)}(t,a)+\sigma^2 f^{(1)}(t,a)+\cdots
\end{equation*}
At the leading order, we obtain the following first order PDE for $f^{(0)}$, which by substitution is:
\begin{equation*}
f^{(0)}_t+\Big(1+\xi\frac{\kappa_t-a}{T-t}\Big)f^{(0)}_a +r(1-\gamma)f^{(0)}-(\rho+\lambda(a))f^{(0)}+\gamma (f^{(0)})^{1-\frac{1}{\gamma}}=0.
\end{equation*}
with the same (as before) boundary condition at $t=T$. This equation, which is obviously simpler than equation \eqref{consumptioneqnfform}, can be solved by using the so-called {\em method of characteristics}, that is by by treating biological age up to first order as a deterministic function $a(t)$ of time $t$ given by:
\begin{equation}
\frac{da}{dt}=1+\xi\frac{\kappa_t-a}{T-t}
\label{characteristics}
\end{equation}
with $a(T)=\kappa_T$, since both ages converge to each other at time $T$. We can then solve (the modified) equation $F(t)=f^{(0)}(t,a(t))$ using the following ordinary differential equation (ODE) representation instead of the original PDE.
\begin{equation*}
\frac{dF}{dt}+r(1-\gamma)F-(\rho+\lambda(a(t)))F+\gamma F^{1-\frac{1}{\gamma}}=0
\end{equation*}
with terminal condition $F(T)=f_T$. Of course, it remains to be seen how {\em good} this analytic approximation is relative to the (numerical) solution to the PDE we reported in the body of the paper. Note in fact that the equation \eqref{characteristics} for $a(t)$ can now be explicitly solved in closed-form as:
\begin{equation*}
a(t)=\kappa_t+(a(0)-\kappa_0)\left(\frac{T-t}{T}\right)^\xi 
\end{equation*}
where $a(0)$ is the biological age at time $0$. The equation for $F$ can be further simplified by introducing $G=F^{\frac{1}{\gamma}}$, for which the equation for $G$ actually becomes linear, leading to: 
\begin{equation*}
\frac{dG}{dt}+1+\frac{1}{\gamma}\left[r(1-\gamma)-(\rho+\lambda(a(t)))\right]G=0.
\end{equation*}
In principle now, this equation can solved explicitly. Note that when both ages match and $a(0)=\kappa_0$, we have $a(t)=\kappa_t$, which is the Gompertz case and $F$ can be expressed by using the incomplete Gamma function, as reported in Milevsky and Huang (2010) for example. Another special case is when $\xi=1$ (and $\sigma=0$), where we have a linear relationship:
 \begin{equation*}
a(t)=\kappa_t+(a(0)-\kappa_0)\left(\frac{T-t}{T}\right)=a(0)+kt,\quad k=\frac{\kappa_0-a(0)}{T}+1.
\end{equation*}
In this (simplified) case, the underlying mortality hazard rate is given by:
\begin{equation*}
\lambda(a(t))=\frac{1}{b}\exp\left(\frac{a(t)-m}{b}\right)=\frac{1}{b}\exp\left(\frac{a(0)+kt-m}{b}\right),
\end{equation*}
which recovers the Gompertz form $\lambda^{\text{G}}_x$, thus leading to a closed-form solution for $f^{(0)}.$ Then, we can improve the accuracy by going to a higher order correction $f^{(1)}$, which by the same logic as before satisfies the following PDE:
\begin{equation*}
f^{(1)}_t+\Big(1+\xi\frac{\kappa_t-a}{T-t}\Big)f^{(1)}_a +r(1-\gamma)f^{(1)}-(\rho+\lambda(a))f^{(1)}+\gamma (f^{(1)})^{1-\frac{1}{\gamma}}=-\frac{f^{(0)}_{aa}}{2}.
\end{equation*}
And, since (in the first approximation) we have already obtained $f^{(0)}$, the right-hand-side is known. The left-hand-side is exactly the same as that for $f^{(0)}$, which again can be solved using the method of characteristics as above. 

\subsection{Distribution of Future Biological Age} 
In this section we derive the (sub) density $g(t,a)\,da = P(A_t\in da, \zeta>t)$, which represents the probability that biological age is within any given `range' $da$ and (of course) that the individual is alive at time $t$ (ie at chronological age $\kappa_t$). The sub-density $g(t,a)$ can then be used to compute all relevant expectations and quantiles, since for any function $\phi$, 
\begin{equation*}
E[\phi(A_t)\mid t<\zeta]=\frac{\int \phi(a) g(t,a)\,da}{\int g(t,a)\,da}.
\end{equation*}
In particular, we can then compute the conditional mean $E[A_t\mid t<\zeta]$ for the biological age at time $t$, the conditional second moment $E[A_t^2\mid t<\zeta]$, and the conditional variance (which recall is $E[A_t^2\mid t<\zeta]-E[A\mid t<\zeta]^2$). The conditional variance, of course, differs from the unconditional expressions worked out earlier.

Recall that by definition aging stops at age $T$ ($=110$ for example), that is, for $t\ge T$ the conditional mean $=\kappa_T$ and the conditional variance is zero. So at this point we are only concerned and work with $t<T$. Let $\phi$ be smooth, with compact support. By It\^o's lemma
\begin{equation*}
\phi(A_t)=\phi(\kappa_0)+\int_0^t \phi'(A_s)\Big(1+\xi\frac{\kappa_s-A_s}{T-s}\Big)\,ds+\int_0^t \phi'(A_s)\,dB_s+\int_0^t \frac{\sigma^2}{2}\phi''(A_s)\,ds.
\end{equation*}
Therefore incorporating the jump, 
\begin{multline*}
\phi(A_t)1_{\{t<\zeta\}}=\phi(\kappa_0)+\int_0^t \phi'(A_s)\Big(1+\xi\frac{\kappa_s-A_s}{T-s}\Big)1_{\{s<\zeta\}}\,ds+\int_0^t \phi'(A_s)1_{\{s<\zeta\}}\,dB_s\\
+\int_0^t \frac{\sigma^2}{2}\phi''(A_s)1_{\{s<\zeta\}}\,ds-\phi(A_\zeta)1_{\{s\ge\zeta\}}.
\end{multline*}
Subtracting the compensated jump to get a continuous martingale, and then taking expectations, we get that
\begin{multline*}
E[\phi(A_t)1_{\{t<\zeta\}}]=\phi(\kappa_0)+\int_0^t E[\phi'(A_s)\Big(1+\xi\frac{\kappa_s-A_s}{T-s}\Big)1_{\{s<\zeta\}}]\,ds\\
+\int_0^t \frac{\sigma^2}{2}E[\phi''(A_s)1_{\{s<\zeta\}}]\,ds-\int_0^tE[\phi(A_s)1_{\{s<\zeta\}}\lambda_s]\,ds.
\end{multline*}
In other words, 
\begin{multline*}
\int \phi(a)g(t,a)\,da=\phi(\kappa_0)+\int_0^t \int \phi'(a)\Big(1+\xi\frac{\kappa_s-a}{T-s}\Big)g(s,a)\,da\,ds\\
+\int_0^t \int \frac{\sigma^2}{2}\phi''(a)g(s,a)\,da\,ds-\int_0^t\int\phi(a)\lambda(a)g(s,a)\,da\,ds.
\end{multline*}
Taking $\frac{\partial}{\partial t}$, we get that 
\begin{multline*}
\int \phi(a)g_t(t,a)\,da=\int \phi'(a)\Big(1+\xi\frac{\kappa_t-a}{T-t}\Big)g(t,a)\,da\\
+\int \frac{\sigma^2}{2}\phi''(a)g(t,a)\,da-\int\phi(a)\lambda(a)g(t,a)\,da.
\end{multline*}
Using integration by parts this becomes
\begin{multline*}
\int \phi(a)g_t(t,a)\,da = \\
\int \phi(a)\Big[-\frac{\partial}{\partial a}\Big(\Big(1+\xi\frac{\kappa_t-a}{T-t}\Big)g(t,a)\Big)
+\frac{\sigma^2}{2}g_{aa}(t,a)-\lambda(a)g(t,a)\Big]\,da.
\end{multline*}
And since $\phi$ was reasonably arbitrary, 
\begin{equation*}
g_t(t,a)=-\frac{\partial}{\partial a}\Big(\Big(1+\xi\frac{\kappa_t-a}{T-t}\Big)g(t,a)\Big)
+\frac{\sigma^2}{2}g_{aa}(t,a)-\lambda(a)g(t,a),
\end{equation*}
with the initial condition being a delta-function.
The boundary conditions are simply that $g=0$ when $a=\pm\infty$.  We have recovered equation (\ref{subg}) as the PDE satisfied  by the (sub) density $g(t,a)$, which is known as the forward equation. This is the PDE we solved (using numerical methods) in section~\ref{NUM} to obtain the distribution (and quantiles) of spending rates at any time $t$ (or age $\kappa_t$).

A special case gives the population survival probability
\begin{equation*}
{}_tp_{\kappa_0}^{\text{pop}}=P(\zeta>t)=\int_{-\infty}^\infty g(t,a)\,da.
\end{equation*}
and therefore the population hazard rate
\begin{equation*}
\lambda_t^{\text pop}=-\frac{1}{{}_tp_{\kappa_0}^{\text{pop}}}\frac{d}{dt}{}_tp_{\kappa_0}^{\text{pop}}
=-\frac{\int_{-\infty}^\infty g_t(t,a)\,da}{\int_{-\infty}^\infty g(t,a)\,da}.
\end{equation*}
Q.E.D.

\end{document}